%% file: main.tex
\documentclass{article}

\usepackage{microtype}
\usepackage{graphicx}
\usepackage{subcaption}
\usepackage{booktabs} % for professional tables
\usepackage{tabularx}
\usepackage{pifont}
\usepackage[most]{tcolorbox}
\usepackage{xcolor}
\usepackage{overpic}
\usepackage{caption}
\usepackage[T1]{fontenc}
\usepackage{lmodern}
\usepackage{array}
\usepackage{longtable} % Allows the table to span multiple pages if needed

\definecolor{mdblue}{RGB}{33,102,172}
\definecolor{mdgreen}{RGB}{34,139,34}
\definecolor{mdred}{RGB}{220,20,60}
\definecolor{framegray}{RGB}{150,150,150}

\tcbset{
  chartcard/.style={
    colback=gray!3,        % <<< 这里就是底色
    colframe=framegray,
    arc=7pt,
    boxrule=0.8pt,
    left=10pt,right=10pt,top=10pt,bottom=10pt
  }
}
\newcommand{\cmark}{\ding{51}} % ✓
\newcommand{\xmark}{\ding{55}} % ✗

\usepackage{hyperref}

\usepackage{enumitem}
\usepackage[accepted]{paper}
\usepackage{amsmath}
\usepackage{amssymb}
\usepackage{mathtools}
\usepackage{amsthm}
\usepackage{tabularx}
\usepackage{booktabs}

\usepackage{booktabs}   % \toprule \midrule \bottomrule \cmidrule
\usepackage{multirow}   % \multirow
\usepackage{colortbl}   % \cellcolor
\usepackage[table]{xcolor}
\usepackage{graphicx}   % \resizebox

\newcommand{\YHY}[1]{{\color{black} #1}}
\usepackage[capitalize,noabbrev]{cleveref}

\theoremstyle{plain}

\theoremstyle{definition}

\theoremstyle{remark}

\usepackage[textsize=tiny]{todonotes}

\icmltitlerunning{ChartWalker: Benchmarking the Cross-Chart RAG Task}

\begin{document}

\twocolumn[
  % \icmltitle{ChartWalker: Learning a Visual Agent to\\
  % Navigate Chart Knowledge Graphs for Cross-Chart Reasoning}
  \icmltitle{ChartWalker: Benchmarking the Cross-Chart RAG Task \\
  with Hierarchical Knowledge Graphs}
  \icmlsetsymbol{equal}{*}
  \icmlsetsymbol{core}{+}
   
  \icmlsetsymbol{corresponding}{$\ddagger$}

  \begin{icmlauthorlist}
    \icmlauthor{Ning Tang}{equal,fdu,comp}
    \icmlauthor{Chenghan Xie}{equal,stf}
    \icmlauthor{Hanyang Yuan}{equal,zju}
    \icmlauthor{Yi Li}{core,zju}
    \icmlauthor{Renhong Huang}{zju}\\
    \icmlauthor{Qian Kou}{comp}
    \icmlauthor{Xiaofeng Shi}{comp,corresponding}
    \icmlauthor{Hua Zhou}{comp}
    \icmlauthor{Jiarong Xu}{fdu,corresponding}
  \end{icmlauthorlist}
  
  \begin{center}
    \vspace{8pt}
    \textsuperscript{1}\textbf{Fudan University}~~~~~~\textsuperscript{2}\textbf{Beijing Academy of Artificial Intelligence}~~~~\textsuperscript{3}\textbf{Stanford University}~~~~~~\textsuperscript{4}\textbf{Zhejiang University}~~~~~~
   \end{center}


  \vskip 0.3in
]
\renewcommand{\thefootnote}{\fnsymbol{footnote}}
\setcounter{footnote}{0}

\footnotetext{\textsuperscript{*}Equal contribution. 
\textsuperscript{+}Core contributors. 
\textsuperscript{$\ddagger$}Corresponding authors.}

\footnotetext{%
\parbox[t]{0.95\linewidth}{\raggedright
Emails: ningtang24@m.fudan.edu.cn, 
jiarongxu@fudan.edu.cn, 
sxf1052566766@163.com, 
\{yuanhanyang,3200105508,renh2\}@zju.edu.cn
}}

\renewcommand{\thefootnote}{\arabic{footnote}}



%\printAffiliationsAndNotice{}  % no special notice (required even if empty)

\input{cpts/0_abstract}

\input{cpts/1_intro}

\input{cpts/6_related_work}

\input{cpts/3_method_new}
\input{cpts/4_agent}

\input{cpts/5_exps}

\input{cpts/8_conclusion}
\section*{Impact Statement}
This paper presents work whose goal is to advance the field of Machine
Learning. There are many potential societal consequences of our work, none
which we feel must be specifically highlighted here.


\bibliography{main}
\bibliographystyle{paper}

%%%%%%%%%%%%%%%%%%%%%%%%%%%%%%%%%%%%%%%%%%%%%%%%%%%%%%%%%%%%%%%%%%%%%%%%%%%%%%%
%%%%%%%%%%%%%%%%%%%%%%%%%%%%%%%%%%%%%%%%%%%%%%%%%%%%%%%%%%%%%%%%%%%%%%%%%%%%%%%
% APPENDIX
%%%%%%%%%%%%%%%%%%%%%%%%%%%%%%%%%%%%%%%%%%%%%%%%%%%%%%%%%%%%%%%%%%%%%%%%%%%%%%%
%%%%%%%%%%%%%%%%%%%%%%%%%%%%%%%%%%%%%%%%%%%%%%%%%%%%%%%%%%%%%%%%%%%%%%%%%%%%%%%
\include{cpts/7_appendix}

%%%%%%%%%%%%%%%%%%%%%%%%%%%%%%%%%%%%%%%%%%%%%%%%%%%%%%%%%%%%%%%%%%%%%%%%%%%%%%%
%%%%%%%%%%%%%%%%%%%%%%%%%%%%%%%%%%%%%%%%%%%%%%%%%%%%%%%%%%%%%%%%%%%%%%%%%%%%%%%

\end{document}

%% file: cpts/0_abstract.tex
\begin{abstract}
Cross-Chart Retrieval-Augmented Generation (RAG) is critical for complex multimodal analysis in various domains. However, existing benchmarks either focus on well-structured tables or generate cross-chart queries via key-point extraction, leading to lexical overlap and logically weak reasoning. To address this, we propose \textbf{ChartWalker}, a novel framework for constructing challenging cross-chart RAG tasks. Specifically, ChartWalker constructs hierarchical knowledge graphs tailored to charts to preserve analytical structure. Furthermore, we employ a structure-aware sampling algorithm to synthesize semantically coherent multi-hop reasoning paths with controllable difficulty and granularity. Based on this framework, we introduce \textbf{ChartWalker-Bench}, a comprehensive benchmark spanning multiple cross-chart query types. Extensive evaluations across representative RAG paradigms reveal significant performance gaps. We further release \textbf{ChartWalker-Agent}, an agentic baseline to support analysis and future system development. Code is available at \url{https://github.com/downing777/ChartWalker_Pub.git}.
\end{abstract}

% Cross-chart retrieval-augmented generation (RAG) is essential for complex multimodal analysis across scientific, business, and political domains. However, existing benchmarks either focus on well-structured tables or generate cross-chart queries via key-point extraction, leading to lexical overlap and logically weak reasoning. We propose ChartWalker, a framework for constructing challenging cross-chart RAG tasks. ChartWalker builds hierarchical chart-specific knowledge graphs to preserve analytical structure and employs a structure-aware sampling algorithm to generate semantically coherent multi-hop reasoning paths with controllable difficulty and granularity. Based on this framework, we introduce ChartWalker-Bench, a diverse benchmark covering multiple domains and cross-chart query types. Extensive evaluations across representative RAG paradigms reveal substantial performance gaps, highlighting the benchmark’s difficulty and diagnostic value. We further release ChartWalker-Agent, an agentic baseline to support analysis and future system development. All data and code are publicly available.

%% file: cpts/1_intro.tex
\section{Inroduction}

%在什么地方讲chart是一个比较复杂的图像数据，信息密度大，而且不像table有明确的结构信息能够被文本化呢？
% Charts are a primary medium for communicating quantitative evidence in science, business, policy, and journalism. Unlike natural images, charts are \emph{information-dense} and \emph{semantically structured}: they encode variables, units, legends, and visual marks that jointly specify relationships and comparisons that are difficult to express succinctly in free-form text. As a result, high-quality chart understanding is not merely a perception problem---it requires recovering structured semantics and supporting downstream analytical operations. 

% Charts are a primary medium for visualizing quantitative statistics in science, business, policy, and journalism. Unlike natural images or tables, they are information-dense and weakly structured. Variables, units, legends, and visual marks are jointly encoded within a chart to specify relationships and comparisons, which are difficult to express succinctly in free-form text. 

%CQA 是很重要的任务
%微调覆盖不了细枝末节的信息 还是需要RAG 检索
\begin{figure}[!t]
    \centering
    \includegraphics[width=\linewidth]{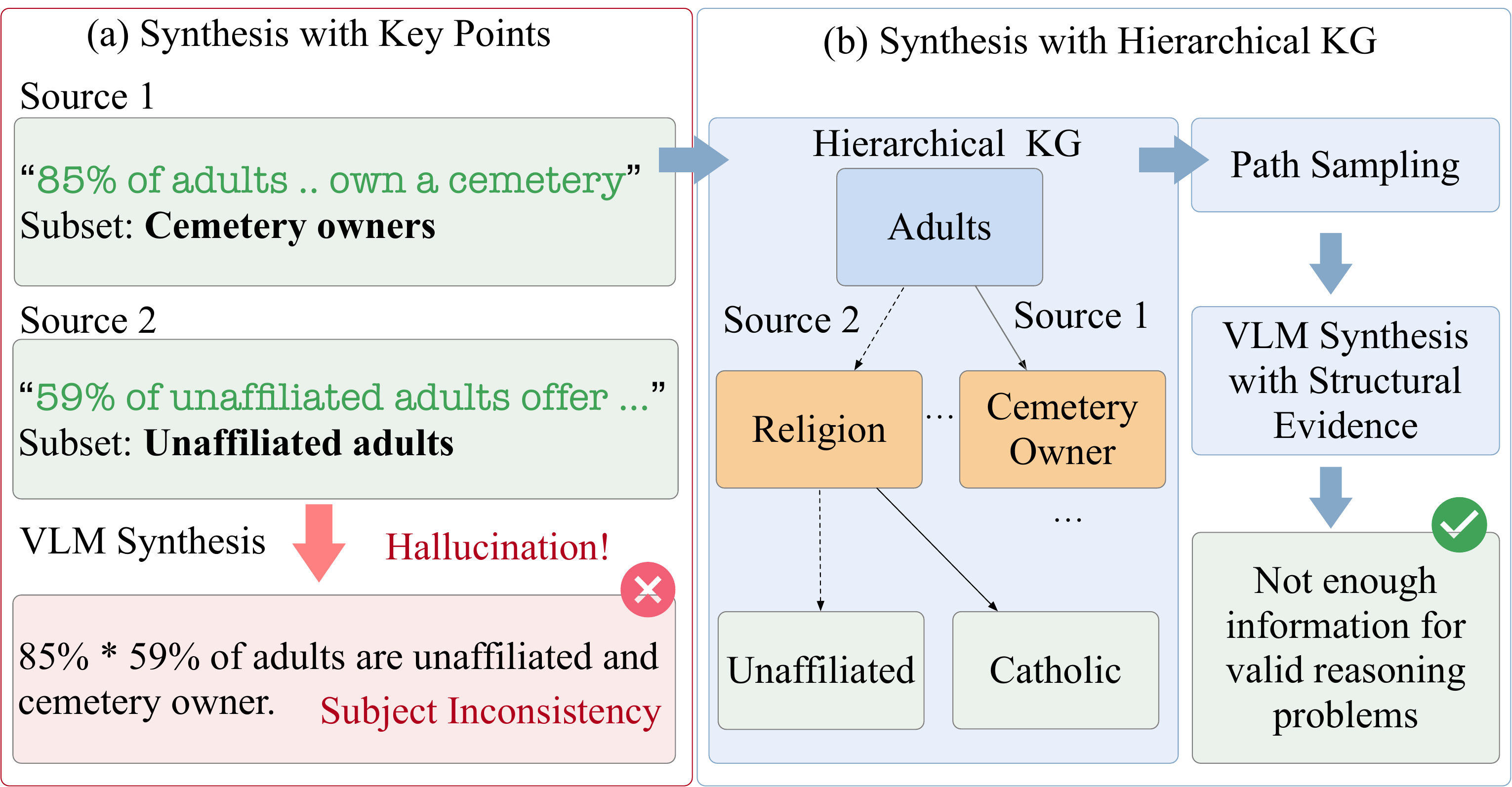}
    \caption{Compared to (a) concatenating isolated key statistics and prompting a VLM to synthesize cross-chart questions, our hierarchical KG (b) explicitly represents entities with their structural relations. Conditioning question generation on these structural paths makes entity dependencies clear and reduces the hallucination of incompatible subjects.}
    \label{fig:subject_mismatch}
    \vspace{-0.2in}
\end{figure}

Charts are a primary medium for visualizing quantitative statistics in science, business, journalism, and policy \cite{DAS2023104896, Norasaed2024, Kastellec_Leoni_2007}. Unlike natural images or tables, they are information-dense and weakly structured. Answering questions by synthesizing evidence across multiple charts is a common requirement in real-world analysis. For instance, an analyst may need to relate a country’s GDP growth trend in one chart to its inflation or unemployment trajectory in another, in order to characterize macroeconomic cycles. 

Recent advances in fundamental Vision Language Models (VLMs) and Multi-modal Language Models (MLMs) \cite{bai2025qwen3vltechnicalreport, openai2024gpt4} have substantially strengthened their capabilities in visual perception and reasoning. When fine-tuned on cross-chart QA questions \cite{masry2025chartqapro}, such models exhibit strong performance in complex numerical and multi-chart analytical tasks. However, fine-tuning alone cannot realistically endow a model with access to all chart-specific knowledge, nor does it generalize well to open-domain or long-tail chart collections encountered in real-world settings. Consequently, many practical applications adopt a Chart Retrieval-Augmented Generation (Chart RAG) paradigm \cite{yang2025benchmarking}, where charts are commonly retrieved as external knowledge sources to support analysis.  Such scenarios are especially common in domains such as scientific surveys, market research, journalism forecasts, and political analysis \cite{kim2020answering, Kastellec_Leoni_2007}.

%Real-world workflows often begin with an underspecified query. To answer this, a system must first locate relevant charts distributed across reports, dashboards, and documents. It then needs to connect evidence from these heterogeneous sources, reconciling potential differences in context, temporal scope, and measurement units before producing a coherent and accurate answer.

% yhy: TODO 这里先说大的再说小的。while this task重要，xxx,但我们发现there lacks a 好的benchmark that具备xxx优点，比如，客观反映真实场景里的query的难度，query本身的逻辑是有意义的（你的优点/你解决了下面这些previous work的什么缺点）。虽然已经有了xxx工作，但他们xxx。因此，我们想做这个事情。
While the Cross-Chart RAG Task is of substantial practical importance, there is a lack of a good benchmark that fully captures the multimodal nature of charts and characterizes the underlying reasoning structure required by realistic queries. We identify two major limitations in existing benchmarks. First, most prior work focuses on tables rather than charts \cite{zou2025ragtableshierarchicalmemory, yu2025tableragretrievalaugmentedgeneration}. In these settings, the underlying data is already explicitly structured, with clear entity boundaries and relations, enabling questions to be constructed and answered through symbolic or textual reasoning alone. 
Moreover, as illustrated in Figure~\ref{fig:subject_mismatch}, recent chart RAG benchmarks \cite{yang2025benchmarking} simply linking semantically similar key points can yield brittle reasoning chains, where implicit referents drift across hops and may lead to subject-mismatch and logically invalid computations (\textit{e.g.}, a follow-up clause refers to a subset while evidence is drawn from the full population).

% 有很多multimodal rag的方法被提出来了，但是没有一个好的evaluation
% 主线 从 mrag出发 kg能提供造QA的工具
%为了揭示一些fundamental的failure case 我做了什么xx

% 把KG的好处和chart 联系起来
% 压缩一下 表达出有粒度和逻辑的问题就可以了
To address these deficiencies, we aim to introduce a logically grounded and complex cross-chart RAG benchmark for evaluating multimodal RAG pipelines. Knowledge graph (KG) provides an intuitive approach to generate such multi-hop QAs: it extracts the implicit entity–relation structure embedded in charts and preserves explicit reasoning paths, which can be directly used to annotate questions with grounded reasoning chains.\cite{lu2025deepdiveadvancingdeepsearch, yang2018hotpotqadatasetdiverseexplainable}. However, existing KG-based QA generation pipelines often rely on random walks or naive PageRank to synthesize long-hop reasoning paths. The path sampling is largely blind to query design, offering limited control over the entity-level constraints that determine granularity and complexity. Moreover, such paths are frequently semantically incoherent: successive hops may be globally unrelated, causing meaningless analysis and sample waste.

To bridge these gaps, we present ChartWalker, a novel chart-centric framework designed to construct challenging cross-chart RAG tasks. Our framework introduces two tightly coupled innovations. First, we propose a hierarchical KG construction method tailored to chart data, which extracts chart entities and relations into explicit layers according to their information granularity. This design enables comprehensive semantic coverage while preserving the inherent structure of dense chart information. Second, building on this hierarchy, we introduce a structure-aware sampling algorithm for cross-chart reasoning path synthesis. Our sampler enforces semantic continuity along paths, ensuring that successive hops remain logically coherent and analytically meaningful. The resulting reasoning paths serve as supervision signals for generating multi-hop QA pairs.

Beyond this methodology, we further release a high-quality cross-chart RAG benchmark, ChartWalker-Bench, comprising 564 multi-hop QA instances across 4 query types. Comprehensive experiments are conducted across major RAG paradigms with different VLM generators. Experiments show that the best-performing model achieves only a 64\% correctness rate in answering the cross-chart problems. More critically, on complex reasoning queries, accuracy drops sharply, falling below 30\% for the majority of cases, highlighting its potential for advancing the multimodal RAG system's capability in multi-step quantitative retrieval and reasoning. In addition, we provide ChartWalker-Agent, a VLM-based search agent baseline that facilitates the analysis of experience reuse and informs future ChartRAG system design. Our main contributions are summarized as follows:
\begin{itemize}[leftmargin=*,parsep=1.1pt,topsep=0pt]
    \item We introduce ChartWalker, a chart-centric framework that explicitly exposes the multi-granular structure of charts by organizing extracted entities and relations into a hierarchical knowledge graph, and synthesizes semantically coherent reasoning paths via structure-aware sampling.
    \item We release ChartWalker-Bench, a curated cross-chart RAG benchmark, with annotations grounded on explicit reasoning chains. Extensive experiments on mainstream RAG baselines show ChartWalker-Bench's difficulty in both retrieval and generation stages.
    \item We further present CharWalker-Agent, a VLM-based search agent for solving the multi-hop reasoning problem, offering insights and experimental analysis for future agent design. 
\end{itemize}
    
\begin{figure*}[!t]
    \centering
    \includegraphics[width=\textwidth]{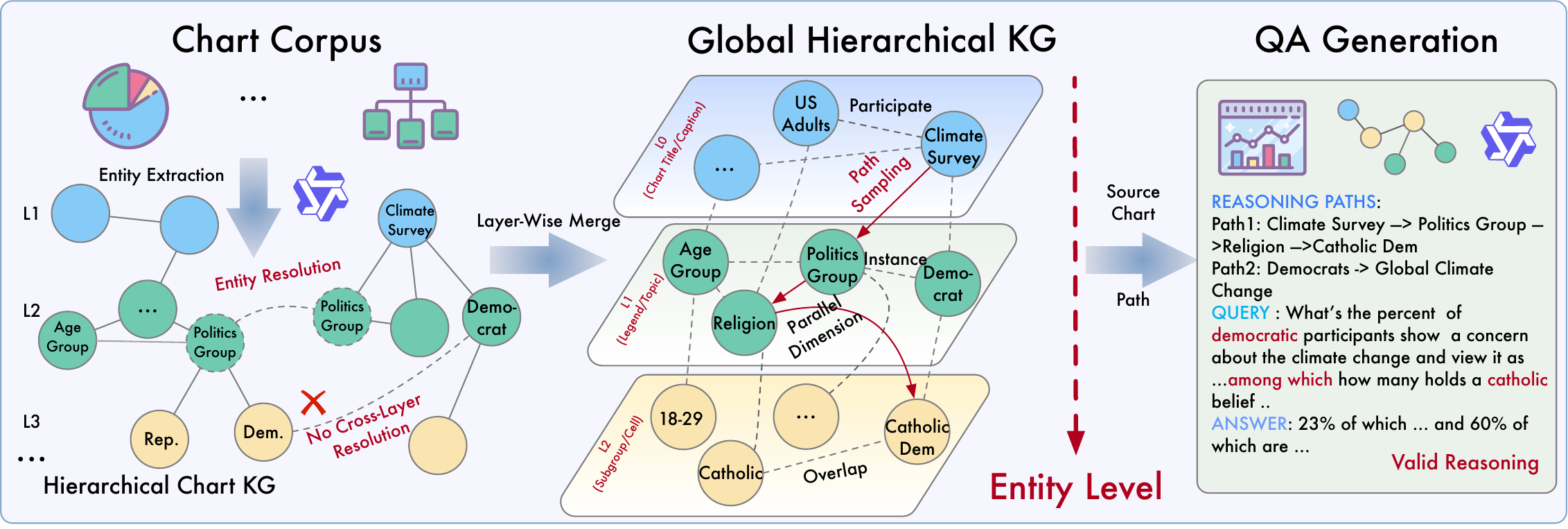}
    \caption{Illustration of the ChartWalker framework. Given a large chart corpus, a VLM first extracts entities and relations from each chart to build per-chart hierarchical knowledge graphs, where entities are organized by granularity levels. Then, identical entities are merged layer-wise to form a global hierarchical KG over the entire corpus. On top of this hierarchy, we perform structure-aware path sampling to construct multi-hop reasoning paths that traverse entities across levels and sources. Finally, we generate cross-chart QA pairs by conditioning on the original chart images and the sampled reasoning paths.}
    \label{fig:chart_walker}
\end{figure*}

\begin{table}[t]
\centering
\caption{Comparison between existing cross-chart benchmarks.}
\label{tab:benchmark_comparison}
\setlength{\tabcolsep}{3.0pt}
\renewcommand{\arraystretch}{1.10}
\scriptsize

\begin{tabularx}{\columnwidth}{l *{4}{>{\centering\arraybackslash}X}}
\toprule
\textbf{Benchmarks} &
\textbf{Objective} &
\textbf{Open Domain} &
\textbf{Structrual Info.} &
\textbf{Complex Reason} \\
\midrule
HeteQA\cite{yu2025tableragretrievalaugmentedgeneration}   &Table           &\cmark  & \cmark &\cmark \\
Open-WikiTable\cite{kweon2023openwikitabledatasetopendomain} &Table      &\cmark  &\xmark &\cmark \\
MTabVQA\cite{singh2025mtabvqaevaluatingmultitabularreasoning} &Table  &\xmark  &\cmark &\xmark \\
MultiTableQA\cite{zou2025ragtableshierarchicalmemory}  &Table           &\cmark  &\xmark &\cmark\\
ChartQA-Pro\cite{masry2025chartqapro}    &Chart       &\xmark  &\xmark &\xmark  \\
Chart-MRAG\cite{yang2025benchmarking}     &Chart   &\cmark  &\xmark &\cmark \\
\midrule
\textbf{ChartWalker(Ours)} &Chart    &\cmark  &\cmark  &\cmark  \\
\bottomrule
\end{tabularx}
\end{table}

%% file: cpts/6_related_work.tex
\section{Related Work}
\paragraph{Chart RAG Benchmark.}
A chart is a general form of visual representation that combines data with graphical marks to convey information. Tables, graphs, and diagrams can all be viewed as instances of charts. Early work focused on
QA over tables, typically assuming a given table context and relatively simple operations\cite{pasupat-liang-2015-compositional, zhong2017seq2sqlgeneratingstructuredqueries}. Research on complex reasoning over charts \cite{li2024mimotablemultiscalespreadsheetbenchmark} emerge with the advance in VLM's capability, where models are required to reason directly over graphical features\cite{masry2025chartqapro} rather than relying on explicit textual or symbolic table schema \cite{xie2025infochartqabenchmarkmultimodalquestion,singh2025mtabvqaevaluatingmultitabularreasoning}. Moving beyond the close domain understanding tasks, \cite{herzig2021opendomainquestionanswering} propose the first research of table RAG problem, where relevant tables must be retrieved from large corpora before reasoning. Building upon this line of work, subsequent studies extended table RAG to more complex cross-table settings \cite{kweon2023openwikitabledatasetopendomain, zou2025ragtableshierarchicalmemory}. The most related work is ChartMRAG \cite{yang2025benchmarking}, which is among the first to benchmark the cross-chart RAG tasks. However, its reliance on semantic similarity yields incoherent reasoning paths, limiting the reliability as an evaluation of cross-chart reasoning. Table \ref{tab:benchmark_comparison} illustrates the main
differences between existing Chart RAG Benchmarks and ChartWalker-Bench.

\vspace{-0.2in}

\paragraph{Multi-hop Question Generation (MHQG).}
MHQG aims to synthesize questions that require multi-step reasoning across multiple contexts \cite{mavi2024multihopquestionanswering}. Early methods largely relied on explicit structured representations to scaffold reasoning.  \cite{kumar2019} formulate difficulty-controllable generation by sampling paths on knowledge graphs, where hop count is used as a proxy for reasoning difficulty.  \cite{liu2023graphsearchnetenhancinggnnscapturing} applies graph convolution to capture global dependencies, enabling evidence aggregation without requiring explicit sentence-level labels.   
Recent studies target both diversity and logical tightness in multi-hop synthesis  \cite{cheng2024multihopquestionansweringtemporal}. KCS \cite{wang-etal-2025-kcs} introduces Knowledge Composition Sampling, which stochastically samples different knowledge combinations from the same context to diversify generated questions while maintaining relevance. In parallel, RT-RAG  \cite{shi2026reasoningtreesimprovingretrievalaugmented} explicitly decomposes complex questions into hierarchical reasoning trees to mitigate error propagation across steps. Despite advances, existing MHQG methods still face a severe scalability challenge, especially in multimodal settings: as hop length increases, these methods suffer from information loss and semantic drift. Consequently, the generated questions often lack meaningful cross-modal grounding and fail to reflect valid multi-hop reasoning.

%% file: cpts/3_method_new.tex
\section{ChartWalker Benchmark}
\label{sec:method}
In this section, we introduce ChartWalker-Bench, a benchmark for chart RAG that provides query--answer pairs with diverse query granularities, along with logically coherent and verifiable rationales that support systematic evaluation. We first formalize the chart retrieval-augmented generation task in \S\ref{subsec:problem}. We then describe our benchmark construction methodology, including: (1) constructing a knowledge graph that explicitly links entities within and across charts (\S\ref{sec:hkg}); and (2) performing path sampling on this graph to obtain coherent multi-hop paths, which are then used to synthesize QA instances with grounded reasoning structures (\S\ref{sec:sampling}). We provide an illustration in Figure \ref{fig:chart_walker}. Finally, we detail how we instantiate this pipeline to construct ChartWalker from real-world data in \S\ref{sec:stats}, including the data sources, processing steps, and resulting dataset statistics.
% \textit{(e.g.}, scale and query-type coverage). 

% This section describes both the resulting dataset and the construction pipeline.
% We first report dataset statistics and positioning against prior benchmarks (\S\ref{sec:stats}), including data sources, scale, query-type coverage, and domain diversity.
% We then detail how the benchmark is generated in two stages: (1) constructing a Knowledge Graph that explicitly links entities within and across charts (\S\ref{sec:hkg}); and (2) performing trajectory sampling on this graph to obtain coherent multi-hop paths, which are then used to synthesize QA instances with grounded reasoning structure (\S\ref{sec:sampling}).

\subsection{Problem Formulation} \label{subsec:problem}
In this work, we focus on the RAG task in the context of charts in their general form, \textit{i.e.}, as images or text-based formats, rather than traditional tables in text form only \cite{kweon2023openwikitabledatasetopendomain,yu2025tableragretrievalaugmentedgeneration}.
% In this work, we focus on the RAG task in the context of charts in their original visual form (\textit{i.e.}, chart images), rather than traditional tables in text form (cite xxx benchmark).
Charts are more ubiquitous in real-world documents and convey information through richer and more flexible visual encodings (\textit{e.g.}, marks, colors, shapes, spatial layouts), making chart RAG both more practical and more challenging to reason over.

\paragraph{Cross-Chart RAG Task}
Let $\mathcal{C}=\{c_j\}_{j=1}^{N}$ be a corpus of charts, 
and let $q$ be a natural-language query.
A ChartRAG system is composed of a retriever $f_{r}$ and a generator $f_{g}$ (\textit{e.g.} a pre-trained VLM). The retriever ranks all charts in $\mathcal{C}$ by relevance to $q$ and returns the top-$k$ charts $\mathcal{C}^k = f_{r}(q;\mathcal{C})$.
% \begin{equation}
% \mathcal{C}_k(q) = f_{r}(q;\mathcal{C}) \subseteq \mathcal{C}, \qquad |\mathcal{C}_k(q)| = k .
% \label{eq:chartrag_retrieval}
% \end{equation}
Conditioned on the query and the retrieved set, the generator produces an answer $\hat{y} = f_{g}\!\left(q, \mathcal{C}^k\right)$. The objective of the cross-chart ChartRAG is to maximize answer correctness.
% \begin{equation}
% \hat{y} = f_{g}\!\left(q, \mathcal{C}_k(q)\right).
% \label{eq:chartrag_generation}
% \end{equation}

% The objective of cross-chart ChartRAG is to maximize answer correctness under a fixed retrieval budget $k$:
% \begin{equation}
% \max \ \mathbb{E}_{(q,y)}\big[ \mathbb{I}\big(\hat{y}=y\big) \big]
% \quad \text{s.t.} \quad |\mathcal{C}_k(q)|=k,
% \label{eq:chartrag_objective}
% \end{equation}
% where $y$ is the ground-truth answer and $\mathbb{I}(\cdot)$ is the indicator function.

Existing evaluations \cite{yang2025benchmarking} on cross-chart RAG can suffer from limitations, as the generated QAs may exhibit subject-mismatch and logically invalid reasoning.
To address this problem, we leverage a knowledge graph to make information links explicit across charts and ultimately generate queries of varying granularity with logically coherent rationales. Specifically, to enable controllable granularity in query generation, we first construct a hierarchical knowledge graph over the entire chart corpus, where entities are organized by their information granularity. To ensure that each query is associated with a logically coherent rationale for the answer, we perform constrained path sampling for the final QA generation. Details are presented below.

\subsection{Hierarchical Chart Knowledge Graph}
\label{sec:hkg}

Inspired by \cite{pasupat-liang-2015-compositional,zhang2020graphrepresentationsemistructureddata}, we seek to represent the chart corpus as a knowledge graph, which provides a uniform substrate for connecting entities across heterogeneous charts and makes multi-hop reasoning explicit. The KG nodes correspond to chart entities (\textit{e.g.}, titles, legends, and individual units), and the edges encode semantic relations.
In practice, as charts convey information at different granularities, user queries can naturally span multiple information scales, such as global context from titles or captions, series- or category-level patterns, and individual unit values; thus, an effective benchmark must account for how RAG methods retrieve evidence across different granularities. However, in a naive KG, entities extracted at different information scales are not explicitly distinguished or organized by granularity \cite{han2025retrievalaugmentedgenerationgraphsgraphrag}. As a result, multi-hop sampling can drift across levels, making the semantic granularity of the resulting QA data difficult to control.
To address this issue, we organize entities into a hierarchical KG that preserves the chart’s inherent information scale. Specifically, we first construct a hierarchical local KG for each chart, and then obtain a unified KG through global integration. This design enables the subsequent generation of multi-hop reasoning paths with controllable information granularity.

\paragraph{Chart Graph Construction.} 
Given a chart $c$, we first prompt a VLM-based extractor to identify structured entities in the chart and annotate each entity with a granularity level. Formally, the extractor returns the entity set as
\begin{equation}
\mathcal{V}_c = \mathsf{VLM}(c, \mathsf{Prompt}_{\mathsf{ent}}),
\end{equation}
where $\mathsf{Prompt}_{\mathsf{ent}}$ denotes the prompt for entity extraction. For $v \in \mathcal{V}_c$, a corresponding granularity level is given as
\begin{equation}
l_v = \mathsf{VLM}(v,\mathsf{Prompt}_{\mathsf{lvl}}) , 
\end{equation}
where $l_v\in \{0,1,\dots,L\}$. Smaller $l$ corresponds to coarser, more globally informative entities (\textit{e.g.}, title entities), and larger $l$ corresponds to finer-grained units (\textit{e.g.}, datapoints). 

{Subsequently, we instantiate edges between extracted entities to capture the relationships between chart components. Intuitively, entities at the same level can exhibit associative relations, while entities at different levels can reflect semantic progression. For example, titles and captions define the topic, axes and legends specify comparison dimensions, and marks and datapoints realize these dimensions with concrete values. We therefore connect two entities with an intra-level edge if they exhibit an associative relation, or with an inter-level edge if they reflect semantic progression.}
Because the same pair of entities can be linked multiple times, the resulting graph is essentially a multigraph and each edge is represented as a relation triple $(v,r,u) \in \mathcal{E}_c$. Together with the extracted entities, we obtain the resulting hierarchical chart KG $\mathcal{G}_c=(\mathcal{V}_c,\mathcal{E}_c)$.

\paragraph{Global Integration.}
The global KG $\mathcal{G}=(\mathcal{V},\mathcal{E})$ is the union of all chart subgraphs:
\begin{equation}
\mathcal{V}=\bigcup_{c\in\mathcal{C}} \mathcal{V}_c,\qquad
\mathcal{E}=\bigcup_{c\in\mathcal{C}} \mathcal{E}_c.
\end{equation}
% \vspace{-0.2in}
The same entity may be mentioned across different charts, and to avoid duplicate entities, we merge identical entities at each level and rewrite edges accordingly. Note that the level does not correspond to a semantic or linguistic hierarchy (\textit{e.g.}, abstract concepts vs. concrete instantiations). Instead, it reflects how informative the entity is to its source chart. Therefore, identical entities can legitimately occur in different layers, and we do not reconcile them across levels.

\subsection{Path Sampling and QA Synthesis}
% 每一步做法的想法
% 补充衔接的语句
% 小标题直接讲具体内容， 例如 anchor selection    
\label{sec:sampling}

As we construct the knowledge graph, the next challenge lies in how to sample paths and use them as supervision signals for QA generation. A central consideration is to ensure that the generated QAs cover diverse information-query granularities and are accompanied by logically coherent rationales. However, existing sampling techniques, such as unconstrained random walks \cite{lu2025deepdiveadvancingdeepsearch} or naive multi-hop expansion \cite{guo2025raganythingallinoneragframework}, may quickly drift to weakly related entities, producing paths that are difficult to convert into coherent, answerable questions, and they also lack a mechanism for granularity-controlled sampling.
% To address this, we impose constraints during sampling to control both the granularity of evidence via level transitions and the semantic coherence around an anchor topic. The sampled trajectories are then converted into grounded QA records with explicit citations.
{To address this, we impose constraints during sampling to control both the granularity of evidence through level transitions and the semantic coherence around an anchor topic. Specifically, before sampling begins, we first select an \textit{anchor} entity as the starting point of the path, considering its importance within the knowledge graph. During the sampling process, we apply constraints to the next-hop sampling policy, which are derived from both semantic topic coherence and information-level considerations. Finally, the sampled paths are converted into grounded QA records with explicit citations.}

\paragraph{Anchor Selection.}

% We start with choosing a starting entity (anchor) that is globally salient and well-connected across the KG. A natural way to do this is to score entities by PageRank\cite{BRIN1998107}, which favors entity nodes that are central in the graph and frequently visited through many high-connectivity paths. However, in the chart KG, an entity node may have a high degree simply because it is repeatedly connected within a single information-dense chart. For example, in a chart that compares multiple countries, the entity \textit{Nation} can be linked to many country-name nodes, yielding a very large local degree; yet these edges all originate from the same chart. These entities are less useful for cross-chart reasoning since they provide limited source diversity. 
We begin by selecting a globally salient and well-connected anchor entity from the KG. A natural approach is to score entities using PageRank \cite{BRIN1998107}, which favors central entities with high connectivity. However, in the chart KG, an entity may have a high degree because it is repeatedly connected within a single, information-dense chart. For example, the entity {``Nation''} may link to many country-name entities in a comparison chart, but these connections lack \textit{source diversity}, making such entities less useful for cross-chart reasoning.

In this sense, we compute the modified PageRank score of entity nodes using a weighted transition probability that jointly considers their connectivity and source diversity. Formally, we define the transition matrix $\boldsymbol{M} = [M_{u,v}]$ as
% \begin{equation}
% M_{u,v} =\frac{w(v,u)}{\sum_{u'\in\mathcal{N}(v)} w(v,u')}
% \end{equation}
\begin{equation}
M_{u,v} = \frac{N_{\mathsf{edge}}(v,u) \cdot N_{\mathsf{src}}(u)}{\sum_{u' \in \mathcal{N}(v)} N_{\mathsf{edge}}(v,u') \cdot N_{\mathsf{src}}(u')},
\end{equation}
where $\mathcal{N}(v)$ denotes neighbors of $v$, $N_{\mathsf{edge}}(v,u)$ is the number of edges between $v$ and $u$, measuring connectivity, and $N_{\mathsf{src}}(u)$ is the number of distinct source charts among all edges related to $u$, representing source diversity. Qualitatively, higher source diversity encourages a stronger transition probability.
% where $\mathcal{N}(v)$ denotes neighbors of $v$ and $w(v,u)$ is defined as:
% \begin{equation}
% w(v,u)=N_{\mathsf{edge}}(v,u)\cdot N_{\mathsf{src}}(u),
% \end{equation}
% where $N_{\mathsf{edge}}(v,u)$ is the number of edges between $v$ and $u$, and $N_{\mathsf{src}}(u)$ is the number of distinct sources among all edges related to $u$. 

The final PageRank score is defined as the stationary distribution $\boldsymbol{\pi}$ of this transition matrix, which satisfies $\boldsymbol{\pi} = \boldsymbol{\pi}\boldsymbol{M}$. During path sampling, the starting entity $v_1$ is then chosen following the distribution $v_1 \sim \boldsymbol{\pi} $.

\paragraph{Path Sampling.}

Since the starting anchor only provides a structural foundation for generating the reasoning path, the quality of the resulting QA depends significantly on how we expand from the anchor to collect supporting evidence. To limit semantic drift and proactively control the query's granularity, we define the next-hop sampling policy \textit{w.r.t} semantic topic coherence and the level of the entities being reached.
Denote a sampled path of $T$ hops as:
% \begin{equation}
% \mathcal{P}_T = \{(v_t, r_t, v_{t+1}|t=1,2,...T, (v_t, r_t v_{t+1}) \in \mathcal{E}\}
% \end{equation}
\begin{equation}
\mathcal{P}^T = \{(v_t, r_t, u_{t}) \mid v_{t+1}=u_t, (v_t, r_t, u_{t}) \in \mathcal{E}\}_{t=1}^{T}.
\end{equation}
% and $\mathcal{P}_t \subset \mathcal{P}_T$ denotes the partial path up to step $t$ (\textit{i.e.} $\mathcal{P}_t = \{(v_1, r_1, v_2),...(v_{t}, r_{t}, v_{t+1})\}$).
In this process, the next hop policy follows:
\begin{equation}
p(v_{t+1},r_{t+1},u_{t+1} \mid \mathcal{P}^t) \propto \pi(v_{t+1}) \cdot \phi_{\text{sem}} \cdot \phi_{\text{gran}}.
\label{eq:branch_policy}
\end{equation}
Here $\phi_{\text{sem}}$ is the cosine similarity between $(v_{t+1},r_{+1},u_{t+1})$ and current path $\mathcal{P}^t$, capturing semantic topic coherence. $\phi_{\text{gran}}$ is a dynamically adjusted scalar function that varies across different sampling processes. It controls the granularity of transitions, assigning different values for same-level and cross-level moves. 
For example, it can favor upward-level transitions (\textit{i.e.} $l_{u_{t+1}} > l_{u_t}$) by assigning larger values, biasing the policy towards higher-level entities, and increasing the query's granularity. Alternatively, the opposite strategy can favor staying at a shallower level to generate coarser-grained queries.
% It is used to control cross-level navigation: when the search prefers moving toward finer-grained evidence, we assign a larger $\phi_{\text{gran}}$ to upward transitions (\textit{i.e.}, when $l_{v_{t+1}} > l_{v_t}$ ), thereby biasing the policy to traverse across levels to reach high level entites
The sampling process terminates when either the maximum hop budget $T$ is reached or the current entity has no outgoing edges.
\paragraph{QA Generation.}
In practice, to generate a high-quality QA instance, we sample multiple paths rooted at the same starting entity, as this provides more avenues to extract information from different perspectives.
% denoted by $\mathcal{B}=\{\mathcal{P}_T^{(i)}\}_{i=1}^{B}$, and collect all source charts appearing on its edges. 
% All this information is packed in a unified prompt skeleton (prompt variants across query types are listed in Appendix~\TN{X}) and a VLM is instructed to output a record containing question, answer, explanation, and explicit evidence usage, while obeying global constraints. 
\YHY{All the sampled paths are packed in a unified prompt skeleton (prompt variants across query types are listed in Appendix~\ref{app:prompt}). A VLM is instructed to formulate questions based on this specific information \textit{w.r.t} different query types (see \S\ref{sec:stats} for detailed query types), and output a complete QA record containing the question, answer, explanation, and explicit evidence usage. We also set constraints where questions must be decontextualized, meaning they are fully self-contained and do not rely on implicit references (\textit{e.g.}, pronouns such as ``this'' or ``that'').}
% In particular, questions must be decontextualized, meaning they are fully self-contained and do not rely on implicit references (\textit{e.g.}, pronouns such as ``this'' or ``that''). 
Additionally, queries exhibiting excessive lexical overlap with the original chart text are paraphrased to reduce direct lexical copying. Formally, we have:
\begin{equation}
\mathsf{QA}=\mathsf{VLM}\left(\{\mathcal{P}_{i}\}_{i=1}^{k}, \mathsf{Prompt}_{\mathsf{gen}} \right),
\end{equation}
where $\{\mathcal{P}_{i}\}_{i=1}^{k}$ denotes the sampled $k$ paths and $\mathsf{Prompt}_{\mathsf{gen}}$ denotes the prompt skeleton for QA generation. This is followed by post-verification to ensure answer correctness against the evidence, with resampling if verification fails.

%Finally, we apply a post-processing step for query decontextualization: questions with high recall-risk are automatically rewritten (e.g., anonymizing overly specific entities/timestamps) to reduce lexical shortcuts while preserving answerability from the provided evidence.

\subsection{Benchmark Construction}
\label{sec:stats}
{Utilizing the proposed construction pipeline, we next introduce how ChartWalker is built upon real-world data.
The original chart corpus is collected from ChartMRAG~\cite{yang2025benchmarking} and ChartQA-pro~\cite{masry2025chartqapro}. Based on this, we further curate the corpus by using a VLM to filter these source charts according to visual clarity, semantic richness, and by merging them based on rule-based duplicate detection.
This process yielded a corpus of 806 charts, encompassing a wide variety of chart styles and in-chart information.}
{Subsequently, we construct a subgraph per chart and merge into a global hierarchical chart KG of 4 layers with 8802 entities and 21436 relations. 
Based on the constructed KG, we perform path sampling with 4 paths per question and up to 4 hops per path, enforcing at least 2 unique chart sources and a maximum of 5 sources per QA pair, yielding 737 generated raw QAs.}
% Based on the built KG, we perform path sampling with 4 paths per question and up to 4 hops per path (enforcing $\ge 2$ unique chart sources ), then generate QA with at most 5 sources, using a reusable path pool of size 120, yielding 737 raw QAs.}
\paragraph{Query types.}
Following the prior work \cite{li2024mimotablemultiscalespreadsheetbenchmark, singh2025mtabvqaevaluatingmultitabularreasoning,zou2025ragtableshierarchicalmemory}, ChartWalker groups the generated queries into 4 types based on common scenarios:
(i) Fact Check,
(ii) Manipulation (sum/average/rank/compare),
(iii) Analysis
and (iv) Complex Reasoning.
% We report the detailed statistics in Table  \ref{tab:benchmark_comparison}. 

\paragraph{Post-verification.}
We apply an automatic quality control step to filter generated QA pairs before constructing the benchmark. The verifier is given the question, the proposed ground-truth answer, and the associated evidence. It outputs strict labels for ``supported'' and ``meaningful'' (each in \{yes, no, uncertain\}). We keep an instance only if both labels are “yes”. The resulting overall pass rate is 0.77. 
% {The resulting pass rates are 0.92 (72/78) for Fact Check, 0.84 (141/168) for Manipulation, 0.83 (242/291) for Analysis, and 0.55 (109/200) for Complex Reasoning, where the lower rate reflects its intentionally harder and more adversarial construction.}
The final filtered result contains 564 QA pairs, which constitute the final ChartWalker. We report key statistics in Table~\ref{tab:extended_query_statistics}.

% \begin{figure}[t]
%     \centering
%     \includegraphics[width=1.0\linewidth]{image/query_type.pdf}
%     \caption{
%     \textbf{Sample distribution across the query types.
%     }}
%     \label{fig:question_count}
% \end{figure}
\paragraph{Reasoning Complexity}
Our benchmark exhibits substantial reasoning complexity. On average, each question draws on 2.30 evidence sources and spans 2.81 reasoning branches among the sampled paths. Notably, the \textit{Complex Reasoning} subset is the most information-intensive, with the highest source charts usage (3.12) and reasoning hops (5.14), creating a challenging setting that stresses a RAG pipeline's ability to retrieve the right charts under distraction and then integrate multiple pieces of evidence into a coherent answer.
To further characterize difficulty, we ask the VLM to provide a subjective difficulty score in $\{1,2,3\}$ for each question. This score is consistent with the statistics above, with {Complex Reasoning} showing the greatest difficulty.

\begin{table}[t]
\centering
\small
\caption{Question statistics by category: \textbf{Src} = average sources; \textbf{Path} = average paths used for query construction; \textbf{Hop} = average reasoning hops; \textbf{Diff} = average subjective difficulty score; \textbf{Pass} = quality control pass rate.}
\label{tab:extended_query_statistics}
\begin{tabular}{lccccccc}
\toprule
\textbf{Category}
& \textbf{Num}
& \textbf{Src}
& \textbf{Path}
& \textbf{Hop}
& \textbf{Diff}
& \textbf{Pass} \\
\midrule
FactCheck
& 72
& 2.03
& 2.44
& 4.04
& 2.00
& 0.92 \\
Manipulation
& 141
& 2.09
& 2.79
& 4.56
& 2.07
& 0.84 \\
Analysis
& 242
& 2.13
& 2.58
& 4.36
& 2.30
& 0.83 \\
Reasoning
& 109
& 3.12
& 3.60
& 5.14
& 3.00
& 0.55 \\
\midrule
Overall
& 564
& 2.30
& 2.81
& 4.52
& 2.34
& 0.77 \\
\bottomrule
\end{tabular}
\vspace{-0.1in}
\end{table}

\begin{table*}[!ht]
\centering
\caption{Retrieval Recall of different RAG baselines on ChartWalker Bench. 
The optimal performance is in bold, and the second-best performance is underlined.}
\label{tab:retrieval_recall}
\setlength{\tabcolsep}{3.0pt}
\small
\begin{tabular}{l l cc  cc  cc  cc  cc}
\toprule
\multirow{2}{*}{Category} 
& \multirow{2}{*}{Methods}
& \multicolumn{2}{c}{Overall}
& \multicolumn{2}{c}{Fact Check}
& \multicolumn{2}{c}{Manipulation}
& \multicolumn{2}{c}{Analysis}
& \multicolumn{2}{c}{Complex Reasoning} \\
\cmidrule(lr){3-4}
\cmidrule(lr){5-6}
\cmidrule(lr){7-8}
\cmidrule(lr){9-10}
\cmidrule(lr){11-12}
& 
& R@5 & R@10
& R@5 & R@10
& R@5 & R@10
& R@5 & R@10
& R@5 & R@10 \\
\midrule

\multirow{3}{*}{Plain RAG}
& BM25  & 37.55  & 49.80  & 37.50  & 57.64  & 53.13  & 66.31  & 38.22  & 51.24  & 15.93  & 20.06  \\
& Text-EmBed & 44.59  & 59.41  & 36.81  & 51.85  & 45.45  & 58.75  & 44.56  & 58.82  & 48.69  & 66.61  \\
& VL-Embed & \textbf{53.90 } & \textbf{71.87 } & \underline{49.07} & \textbf{73.15 } & \textbf{61.82 } & \textbf{79.91 } & \textbf{52.00 } & \underline{70.52} & \textbf{51.09 } & \underline{63.64} \\

\midrule

\multirow{3}{*}{Graph-based RAG}
& RA-Local & 40.19  & 50.51  & 37.27  & 48.38  & 47.46  & 58.75  & 44.28  & 52.69  & 23.61  & 36.44  \\
& RA-Mix   & 42.38  & 54.19  & 33.10  & 45.60  & 42.97  & 54.26  & 45.11  & 54.13  & 41.68  & 59.89  \\
& HippoRAG & \underline{53.25} & \underline{71.78} & \textbf{53.24 } & \underline{71.76} & \underline{60.75} & \underline{76.00} & \underline{50.28} & \textbf{70.80 } & \underline{50.15} & \textbf{68.52 } \\ 

\bottomrule
\end{tabular}
\vspace{-0.1in}
\end{table*}

%% file: cpts/4_agent.tex
\section{ChartWalker Agent}
\label{sec:agent}

% \YHY{why agentic necessary for chart rag: describe \& add citation}
% 是一个比较难的任务 表很复杂，跨表又需要推理能力XX
% 非agentic的方法比较能均衡的达到好的效果 看实验结果
% for extension, 一个未来promising 的方向是agentic 为什么是有帮助的 有记忆，动态检索

% 优点和chart 联系起来
% design 的目的是什么 provide insight
% 用absolute value 例如complex reasoning agent 能解决 long hop reasoning 
% A central challenge of cross-chart reasoning lies not merely in chart perception itself, but in the heterogeneity of reasoning demands across query types. Different questions require evidence to be retrieved at different granularities and combined through distinct reasoning strategies, making a single static retrieval pipeline fundamentally insufficient to consistently perform well across diverse reasoning types \TN{ref}.

% As shown in Table~\ref{tab:retrieval_recall}, existing retrieval-based approaches exhibit a pronounced performance degradation on complex multi-hop reasoning queries: even the strongest model achieves only \TN{51\%} answer accuracy with 10 retrieved source charts. In most cases, the accuracy remains below 30\% and indicates that a single static rag pipeline may be insufficient for the long reasoning chains and multi-granularity evidence retrieval across diverse cross-chart reasoning queries.
{A key challenge in cross-chart reasoning lies beyond chart perception, stemming from the heterogeneous reasoning requirements across query types. Different queries demand evidence to be retrieved at varying levels of granularity and composed through distinct reasoning processes, which limits the effectiveness of static retrieval pipelines.
As demonstrated in our experiments (Table~\ref{tab:retrieval_recall}), existing static retrieval-based approaches exhibit notable performance degradation on complex multi-hop reasoning queries. Even with up to 10 retrieved source charts, the strongest model attains only 51\% answer accuracy. In most settings, performance remains below 30\%. These results suggest that a single static RAG pipeline is insufficient to support long-horizon reasoning and multi-granularity evidence aggregation across diverse cross-chart queries.}

Recent advances in agentic retrieval-augmented generation have shown that treating retrieval as a sequential decision-making process can substantially improve long-horizon reasoning \cite{singh2025agenticretrievalaugmentedgenerationsurvey}.
By allowing models to iteratively observe the source corpus, maintain searching memory, and adapt a dynamic retrieval strategy, agent-based approaches offer a promising mechanism for handling the complex tasks.
{Motivated by these developments \cite{geng2025webwatcherbreakingnewfrontier, lu2025deepdiveadvancingdeepsearch}, we design the ChartWalker Agent, which serves as a baseline for better benchmarking cross-chart long-horizon reasoning. The ChartWalker agent is a VLM-based search agent that navigates a chart knowledge graph to iteratively acquire evidence and perform step-wise reasoning, functioning as a diagnostic and analytical tool for studying cross-chart RAG behaviors and informing future research.}

%For fairness, we use the general KG (as constructed in RagAnything \YHY{cite}) as the interactive environment. 

\paragraph{Environment and Action Space.}
% ChartWalker agent is asked to operate on a KG constructed from chart entities and their relations. \YHY{To prevent potential information leakage from reusing the hierarchical KG built during benchmark construction, we follow \cite{guo2025raganythingallinoneragframework} to rebuild a standard KG. In this process, we generate a summary for each chart and treat the summaries as textual passages for KG construction. The agent's goal is to iteratively acquire evidence by exploring the KG, retrieving relevant chart entities and source charts, and terminating once sufficient evidence has been collected. 
The ChartWalker agent operates on a KG constructed from chart entities and their relations. {To prevent potential information leakage from reusing the hierarchical KG built during benchmark construction, we follow \cite{guo2025raganythingallinoneragframework} to rebuild a standard KG. In this process, we generate a summary for each chart and treat the summaries as textual passages for KG construction.} Accordingly, the agent aims to iteratively acquire evidence by exploring the KG, retrieving relevant chart entities and source chart information, and terminating once sufficient evidence has been collected.
To avoid context overflow and unstable credit assignment caused by concatenating long interaction trajectories, we formulate the problem as a partially observable Markov decision process (POMDP) and compress the agent's memory into the environment observation. Specifically, the observation provided to the agent summarizes the current search state, including the current entity, its neighboring entities, and relations in the KG, as well as a simplified record of past search history with retrieved source charts.
At each turn, the agent decides to act from $\{\mathsf{move}, \mathsf{edge\_search}, \mathsf{backward},  \mathsf{stop}\}$. See more environment and action design at Appendix~\ref{app:agent}.

\paragraph{Training Objective and Optimization.}
During exploration, the agent observes the current KG context along with retrieved evidence, takes an action to further explore the KG, and receives a scalar reward from the environment at each turn. The training objective is to maximize the expected discounted return \cite{schulman2017proximalpolicyoptimizationalgorithms}.
In our setting, the agent follows a policy parameterized by a VLM, which autoregressively outputs a token sequence as the action. Following~\cite{wang2025vagen}, we adopt non-concatenated PPO rollouts and a turn-level advantage assignment scheme. Specifically, we estimate the advantage function using temporal-difference learning and assign the resulting turn-level advantage uniformly to all action tokens within each turn. Implementation details are deferred to Appendix ~\ref{app:agent}. We evaluate the performance of the ChartWalker agent and present analyses and insights on the cross-chart RAG task in \S\ref{subsec:gen_perfom}.

% Formally, at turn $t$, the agent receives an observation $o_t$ (the current KG context and retrieved evidence), generates an action $a_t$ to further explore the KG, and the environment returns a scalar reward $r_t$. The objective is to maximize the expected discounted return $\max_\theta \mathbb{E}_{\pi_\theta}\big[\sum_{t=0}^{T-1}\gamma^t r_t\big]$, where $\gamma$ is a discount factor. 
% In our setting, the policy $\pi_\theta$ is parameterized by a VLM that autoregressively outputs a token sequence as the action. Following ~\cite{wang2025vagen}, we adopt a non-concatenated PPO rollout and corresponding advantage estimation. 
% At each turn $t$, the policy conditions the system prompt and current observation $s_t = [\text{sys}, o_t] $
% and autoregressively generates a textual action as a token sequence
% $a_t$.
% Accordingly, the policy ratio can be computed per turn without requiring a forward pass over the entire concatenated trajectory. Let
% \begin{equation}
% \begin{aligned}
% J_{\text{PPO}}(\theta)
% =
% \frac{1}{T}
% \sum_{t=0}^{T-1}
% \min\Big(
% u_t(\theta) A_t,\;
% \mathrm{clip}\!\left(u_t(\theta), 1-\epsilon, 1+\epsilon\right) A_t
% \Big),
% \end{aligned}
% \end{equation}
% We estimate the advantage function $A_t$ using temporal-difference (TD) learning, and assign the same turn-level advantage to all action tokens within turn $t$. We leave the detailed algorithm at \TN{Appendix}.

%% file: cpts/5_exps.tex
\section{Experiments}
\vspace{-0.1in}
% 导言区需: \usepackage{booktabs}  % 可选: \usepackage[tableposition]{caption}

\begin{table*}[t]
  \centering
  \small
  \caption{Entity extraction quality judged against chart images (\textsc{VLM Judge} = \texttt{gpt-5.4}). Metrics are averaged over extracted entities. $N_{\mathrm{ent}}$: total extracted entities evaluated per dataset.}
  \label{tab:entity-eval}
  \begin{tabular}{lcccccc}
    \toprule
    \textbf{Dataset} &
    $\mathbf{P}$ &
    $\mathbf{R}$ &
    \textbf{Halluc.} &
    \textbf{Lvl.} &
    $N_{\mathrm{ent}}$ &
    \textbf{Parse err.} \\
    \midrule
    ChartQA   & 0.587 & 0.606 & 0.185 & 0.655 & 2{,}709 & 0 \\
    ChartMRAG & 0.501 & 0.427 & 0.094 & 0.613 & 1{,}730 & 1 \\
    \bottomrule
  \end{tabular}\\[0.35em]
  \footnotesize
  \textit{Notes.} \textbf{Halluc.}\,=\,micro hallucination rate; \textbf{Lvl.}\,=\,micro level (granularity) accuracy.
  Macro chart-mean: ChartQA ($P{=}0.633$, $R{=}0.593$, Halluc.${=}0.109$, Lvl.${=}0.649$);
  ChartMRAG ($P{=}0.495$, $R{=}0.425$, Halluc.${=}0.089$, Lvl.${=}0.602$).
  ChartQA entity vs.\ relation runs use \emph{different} 100-chart samples (see text).
\end{table*}

\begin{table*}[t]
  \centering
  \small
  \caption{Relation triple quality (same judge). \textbf{Micro} metrics over evaluated triples; at most 40 relations sampled per chart. $N_{\mathrm{rel}}$: total relation judgements.}
  \label{tab:relation-eval}
  \begin{tabular}{lcccccc}
    \toprule
    \textbf{Dataset} &
    $\mathbf{P}$ &
    $\mathbf{R}$ &
    \textbf{Halluc.} &
    \textbf{Type OK} &
    $N_{\mathrm{rel}}$ &
    \textbf{Parse err.} \\
    \midrule
    ChartQA   & 0.647 & 0.609 & 0.082 & 0.748 & 2{,}595 & 1 \\
    ChartMRAG & 0.620 & 0.542 & 0.092 & 0.751 & 2{,}288 & 1 \\
    \bottomrule
  \end{tabular}\\[0.35em]
  \footnotesize
  \textit{Notes.} \textbf{Type OK}\,=\,micro type appropriateness.
  Macro chart-mean: ChartQA ($P{=}0.631$, $R{=}0.576$, Halluc.${=}0.082$, Type${=}0.734$);
  ChartMRAG ($P{=}0.619$, $R{=}0.516$, Halluc.${=}0.082$, Type${=}0.741$).
\end{table*}
\begin{table*}[htbp]
\centering
\small
\caption{Answer correctness under different VLM generators on ChartWalker Bench.}
\label{tab:success_rate_vlm}
\setlength{\tabcolsep}{3.0pt}
\resizebox{\textwidth}{!}{%
\begin{tabular}{ll cc  cc  cc  cc  cc}
\toprule
\multirow{2}{*}{VLM}
& \multirow{2}{*}{Methods}
& \multicolumn{2}{c}{Overall}
& \multicolumn{2}{c}{FactCheck}
& \multicolumn{2}{c}{Manipulation}
& \multicolumn{2}{c}{Analysis}
& \multicolumn{2}{c}{Complex Reasoning} \\
\cmidrule(lr){3-4}
\cmidrule(lr){5-6}
\cmidrule(lr){7-8}
\cmidrule(lr){9-10}
\cmidrule(lr){11-12}
& 
& Cor@5 & Cor@10
& Cor@5 & Cor@10
& Cor@5 & Cor@10
& Cor@5 & Cor@10
& Cor@5 & Cor@10 \\
\midrule
\multirow{4}{*}{Qwen3-VL-8B}
& BM25  & 18.44  & 22.69  & 25.00  & 33.33  & 15.60  & 18.44  & 25.62  & 29.75  & 1.83  & 5.50  \\
          & Text-EmBediing & 26.59  & 35.46  & 36.11  & 41.67  & 24.11  & 32.62  & 33.05  & 44.63  & 9.17  & 14.68  \\
          & VL-Embedding & \underline{42.02} & \underline{49.65} & \underline{59.72} & \underline{72.22} & \underline{44.68} & \underline{56.03} & \underline{48.76} & \underline{55.37} & 11.93  & 13.76  \\  
          & Local & 27.84  & 35.11  & 37.50  & 50.00  & 24.11  & 29.79  & 36.77  & 41.74  & 6.42  & \underline{17.43} \\
          & Mix   & 37.94  & 41.14  & 47.22  & 48.61  & 38.30  & 42.55  & 46.28  & 48.76  & \underline{12.84} & \underline{17.43} \\
          & HippoRAG & \textbf{45.74 } & \textbf{59.40 } & \textbf{56.94 } & \textbf{76.39 } & \textbf{46.10 } & \textbf{60.28 } & \textbf{51.65 } & \textbf{63.64 } & \textbf{24.77 } & \textbf{37.61 } \\
\midrule
\multirow{4}{*}{Qwen3-VL-32B}
& BM25  & 20.74  & 25.36  & 22.22  & 38.88  & 17.02  & 22.70  & 27.27  & 29.34  & 10.09  & 11.01  \\
          & Text-EmBediing & 26.59  & 37.06  & 34.72  & 45.83  & 19.15  & 34.04  & 32.23  & 44.62  & 18.35  & 18.35  \\
          & VL-Embedding & \underline{48.58} & \underline{59.4} & \underline{63.89} & \underline{79.17} & \underline{50.35} & \textbf{68.79 } & \textbf{54.13 } & \underline{62.81} & \underline{23.85} & \underline{26.61} \\
          & Local & 28.37  & 37.05  & 38.88  & 51.39  & 23.40  & 33.33  & 36.36  & 44.21  & 10.09  & 16.51  \\
          & Mix   & 40.42  & 45.39  & 55.55  & 55.55  & 36.17  & 41.13  & 48.76  & 53.72  & 17.43  & 25.69  \\
          & HippoRAG & \textbf{50.18 } & \textbf{64.72 } & \textbf{68.06 } & \textbf{77.78 } & \textbf{51.06 } & \underline{66.67} & \underline{52.07} & \textbf{68.60 } & \textbf{33.03 } & \textbf{44.95 } \\
\midrule
\multirow{4}{*}{GPT-4o}
 & BM25  & 29.79  & 33.87  & 38.88  & 47.22  & 22.69  & 22.70  & 36.36  & 38.43  & 18.35  & 29.36  \\
          & Text-EmBediing & 26.77  & 36.52  & 31.94  & 40.28  & 22.69  & 30.50  & 32.23  & 44.21  & 16.51  & 24.77  \\
          & VL-Embedding & \textbf{56.03 } & \textbf{64.89 } & \textbf{66.67 } & \textbf{76.39 } & \textbf{49.65 } & \textbf{66.67 } & \textbf{61.57 } & \underline{67.36} & \textbf{44.95 } & \underline{49.54} \\
          & Local & 27.68  & 35.44  & 31.94  & 43.86  & 24.29  & 34.48  & 36.40  & 45.77  & 10.09  & 10.28  \\
          & Mix   & 42.93  & 46.31  & 54.29  & 66.67  & 35.00  & 39.47  & 54.58  & 57.45  & 20.18  & 27.52  \\
          & HippoRAG & \underline{49.82} & \underline{63.30} & \underline{59.72} & \underline{70.83} & \underline{46.81} & \underline{56.03} & \underline{55.37} & \textbf{70.66 } & \underline{34.86} & \textbf{51.38 } \\

\bottomrule
\end{tabular}}
\vspace{-0.1in}
\end{table*}
We evaluate (i) the retrieval quality of different RAG pipelines and (ii) the effectiveness of VLM generators in leveraging retrieved sources for correct answering.

\vspace{-0.1in}
\subsection{Experimental Setup}

\paragraph{Baselines.}
We compare against five representative retrievers, grouped into two families: plain RAG (by similarity) and graph-based RAG. Plain RAG involves BM25 \cite{SIGIR-1994-RobertsonW}, Dense textual embedding \cite{qwen3embedding} and Vision-Language embedding \cite{qwen3vlembedding}; while graph-based RAG involves RagAnything (local/mix) \cite{guo2025raganythingallinoneragframework} and  HippoRAG \cite{gutiérrez2025ragmemorynonparametriccontinual}. 

For VLM generators, we test three VLMs with different scales and openness: Qwen3-VL-8B, Qwen3-VL-32B \cite{bai2025qwen3vltechnicalreport}, and GPT-4o\cite{openai2024gpt4}. Unless otherwise stated, we keep the prompting format fixed across the models to isolate the effect of model capability.

\paragraph{Evaluation Metrics}
\label{sec:metrics}

To assess retrieval quality, we report Recall@$k$ (R@$k$), which measures the fraction of annotated gold evidence sources that appear in the top-$k$ retrieved list. To evaluate end-to-end success, we report Correctness@$k$ (Cor@$k$), defined as the proportion of queries that the VLM generators answer correctly when conditioned on the top-$k$ retrieved sources. Correctness is evaluated using an LLM-as-a-judge paradigm \cite{zheng2023judgingllmasajudgemtbenchchatbot}.
Unless otherwise specified, we report results at $k=5$ and $10$, and all metrics are reported as percentages (\%).

\paragraph{Implementation Details.}
For the text-based RAG framework, we use Qwen3-VL-8B-Instruct ~\cite{bai2025qwen3vltechnicalreport} to extract structured summaries and entities from each chart to build the multimodal database and knowledge graph.  Dense embeddings are generated by Qwen3-Embedding-8B \cite{qwen3embedding}  and  Qwen3-VL-Embedding-8B \cite{qwen3vlembedding}. All generator models use a fixed zero-shot prompt template and greedy decoding (temperature=0.0) for deterministic outputs. For agent training and correctness comparison \ref{tab:base_vs_agent}, we use Qwen2.5-VL-3B-Instruct \cite{bai2025qwen25vltechnicalreport} as the base policy model and VLM generator. A more detailed description and parameter settings of baselines can be found in Appendix~\ref{app:exp}.

All experiments are conducted on a machine with Ubuntu 22.04 system, equipped with AMD EPYC 7742 64-Core Processor and 8× NVIDIA A100 GPUs (40GB memory). VL-Embedding is implemented in PyTorch version 2.8.0 with CUDA version 12.8 and Python 3.13.11, and others are all implemented in PyTorch version 2.9.1 with CUDA version 12.8 and Python 3.10.19.

\subsection{Retrieval Performance}
Table~\ref{tab:retrieval_recall} reveals that retrieval remains far from saturated (best R@10 $\approx$ 72), confirming that ChartWalker-Bench is non-trivial under a limited top-$k$ budget. 

{\bf Enhanced multimodal representations outweigh complex retrieval heuristics.} The table shows that the VL-Embedding retriever achieves the strongest average recall among all methods (53.90/71.87), and ranks first across most categories. This suggests that directly aligning query text with chart visuals in a shared embedding space substantially strengthens relevance matching.

{\bf Graph-aware retrieval provides clear benefits.} HippoRAG consistently outperforms the text-only retriever by a large margin (Overall: +8.66 R@5 and +12.37 R@10), with especially pronounced gains on Fact Check and Manipulation, indicating that propagation over the knowledge graph effectively aggregates multi-hop evidence and recovers missing supporting sources that a single-pass embedding search tends to miss. Moreover, HippoRAG is particularly competitive on the more adversarial regimes: it attains the best R@10 on Analysis and Complex Reasoning, aligning with the intuition that graph-based relevance diffusion is helpful when queries are de-lexicalized and evidence is scattered across multiple charts. 
% In contrast, naive neighborhood expansion (RA-Local/RA-Mix) lags behind, highlighting that graph structure alone is insufficient—how the model propagates and consolidates evidence matters.

\vspace{-0.1in}

\subsection{Generation Performance}
\label{subsec:gen_perfom}
Table~\ref{tab:success_rate_vlm} reports answer correctness under different retrievers and VLM generators. Overall, ChartWalker-Bench remains challenging at the generation stage: even with the strongest configuration, overall Cor@10 peaks at $\sim$65, while Complex Reasoning is consistently the hardest subset (best Cor@10 $\approx$ 51), indicating that multi-source retrieval and evidence composition are still major bottlenecks. 

{\bf Stronger retrieval quality translates into substantial gains in correctness, but recall alone does not fully predict success.} For Qwen3-VL-8B, moving from BM25 to dense/multimodal retrieval yields large jumps, and HippoRAG further improves to 45.74/59.40. A similar pattern holds for Qwen3-VL-32B. These results suggest that HippoRAG’s graph-based relevance propagation provides more than “extra hits”: even when its retrieval recall is not the highest, it tends to surface structurally connected evidence that completes multi-hop chains, which is easier for the generator to integrate and reason over the information.

{\bf Scaling generator improves both accuracy and robustness.} Qwen3-VL-32B consistently outperforms Qwen3-VL-8B under the same retriever, and GPT-4o paired with VL-Embedding achieves the best overall scores, showing strong synergy between high-capacity VLMs and unified vision-language retrieval. Interestingly, even with GPT-4o, graph-based retrieval remains valuable for the most challenging reasoning regimes: HippoRAG becomes best at Cor@10 on Analysis (70.66) and Complex Reasoning (51.38), suggesting that structured, multi-hop evidence retrieval complements stronger generative reasoning rather than being replaced by it.

\subsection{Agent Performance and Analysis}
We evaluate ChartWalker-Agent on ChartWalker-Bench with an 8:2 split for training/testing, resulting in a held-out test set of 105 questions (Table~\ref{tab:base_vs_agent}).
For lightweight VLMs (3B), increasing the retrieval budget does not monotonically improve correctness: feeding more charts quickly runs into multi-image context/visual token limitations and introduces distractors, so $k{=}10$ can be worse than $k{=}5$ (e.g., HippoRAG drops from 31.74 at $k{=}5$ to 29.79 at $k{=}10$).
After PPO training, the agentic policy improves evidence acquisition via deeper KG exploration and achieves higher overall accuracy than static pipelines (33.14 vs. 31.74 best static), with the largest gains on the most search-intensive subset, Complex Reasoning.
% We also observe consistent improvements on \textit{Manipulation}, while \textit{FactCheck} remains relatively less benefited, suggesting that verification-style queries are often bottlenecked by precise first-hop chart grounding rather than long-horizon exploration.

\vspace{-0.1in}
\begin{table}[!ht]
\centering
\caption{Answer correctness comparison between HippoRAG and Agent on ChartWalker Bench.}
\label{tab:base_vs_agent}
\setlength{\tabcolsep}{3.0pt}
\renewcommand{\arraystretch}{1.10}
\scriptsize

\begin{tabular}{lccccc}
\toprule
\textbf{Method} & \textbf{Overall} & \textbf{FactCheck} & \textbf{Manipulation} & \textbf{Analysis} & \textbf{Complex} \\
\midrule
VL-Emb(k=5) & 23.76  &34.72  &20.56  &28.51  & 10.09 \\
VL-Emb(k=10) & 25.89  &37.50  &20.57 &32.23  & 11.01 \\
HippoRAG(k=5) & 31.74  &44.44  &29.79  &36.36  & 15.60 \\
HippoRAG(k=10) & 29.79 &37.50  &29.79  &34.30  &14.68  \\
Agent(3b)    &33.14      & 37.50      &  34.37     & 34.08     & 18.75      \\
\bottomrule
\end{tabular}
\end{table}

%% file: cpts/8_conclusion.tex
\vspace{-0.1in}
\section{Conclusion}
In this paper, we introduced ChartWalker, a novel framework and benchmark for Cross-Chart RAG. By leveraging hierarchical knowledge graphs and structure-aware sampling, ChartWalker generates complex, multi-hop reasoning paths that challenge existing systems. Our evaluations on ChartWalker-Bench reveal that current Vision-Language Models struggle with multi-chart analysis, exposing the limitations of static retrieval. To bridge this gap, we proposed ChartWalker-Agent, demonstrating the power of iterative, graph-based evidence acquisition. This work provides a rigorous foundation for advancing multimodal RAG, with future efforts directed toward enhancing multimodal embeddings and agentic reasoning for complex chart analysis.

%% file: cpts/7_appendix.tex
\newpage
\appendix
\onecolumn
\section{Appendix.}
\subsection{Notations}\label{subsec: notation}
\renewcommand{\arraystretch}{1.3} % Increases row spacing for readability

\begin{longtable}{@{} p{0.15\textwidth} p{0.80\textwidth} @{}}
\caption{Summary of Notations and Symbols used in ChartWalker.} \label{tab:notation} \\
\toprule
\textbf{Symbol} & \textbf{Description} \\
\midrule
\endfirsthead

\multicolumn{2}{c}%
{{\bfseries \tablename\ \thetable{} -- continued from previous page}} \\
\toprule
\textbf{Symbol} & \textbf{Description} \\
\midrule
\endhead

\midrule
\multicolumn{2}{r}{{Continued on next page}} \\
\bottomrule
\endfoot

\bottomrule
\endlastfoot

% --- Section 3.1: Problem Formulation ---
\multicolumn{2}{l}{\textit{\textbf{Problem Formulation}}} \\
$\mathcal{C}, c$ & The corpus of charts, chart. \\
$q$ & Natural language query\\
$\mathcal{C}_k()$ & The set of top-$k$ retrieved charts.  \\
$y,\hat{y}$ & The ground-truth answer and The predicted answer generated by the reader model. \\

% --- Section 3.2: Hierarchical KG ---
\midrule
\multicolumn{2}{l}{\textit{\textbf{Hierarchical Knowledge Graph}}} \\
$\mathcal{G}_c, \mathcal{V}_c, \mathcal{E}_c$ & The chart hierarchical knowledge graph, Entities and Edges for chart $c$ \\
$l$ & Granularity level of an entity. \\
$\mathcal{G}$ & The global knowledge graph integrated from all chart subgraphs. \\

% --- Section 3.3: Trajectory Sampling ---
\midrule
\multicolumn{2}{l}{\textit{\textbf{Path Sampling \& QA Synthesis}}} \\
$M$ & The transition matrix for PageRank calculation, $M=[M_{u,v}]$. \\
$\mathcal{N}(v)$ & The set of neighbor entities of entity $v$. \\
$N_{src}(u)$ & Number of distinct source charts associated with entity $u$. \\
$\pi$ & The stationary distribution (PageRank score) used for anchor selection. \\
$\mathcal{P}_T$ & A sampled reasoning path of length $T$. \\
$\phi_{sem}$ & Semantic topic coherence function (cosine similarity). \\
$\phi_{gran}$ & Granularity control function for regulating level transitions. \\

% --- Section 4: ChartWalker Agent ---
% \midrule
% \multicolumn{2}{l}{\textit{\textbf{ChartWalker Agent (RL)}}} \\
% $s_t$ & The state/context at turn $t$. \\
% $o_t$ & The observation at turn $t$ (current KG context and retrieved evidence). \\
% $a_t$ & The action generated by the agent at turn $t$ (token sequence). \\
% $r_t$ & The scalar reward received from the environment at turn $t$. \\
% $\pi_\theta$ & The policy parameterized by a VLM. \\
% $V_\phi(s_t),A_t, G_t $ & The value function (critic) defined on turn context; The advantage function at turn $t$ and The return target for the value function. \\

\end{longtable}
\subsection{More experiment settings}
\label{app:exp}

\paragraph{Baseline Details} 
(1) BM25 \cite{SIGIR-1994-RobertsonW}: a sparse lexical retriever that ranks candidates by term-matching scores. (2) Dense textual embedding \cite{qwen3embedding}: we convert each candidate visual source into text, embed both query and candidates, and rank by cosine similarity. (3)Vision-Language embedding \cite{qwen3vlembedding}: unify textual and visual features into one embedding space and compute the direct similarity. (4) RagAnything (local/mix) \cite{guo2025raganythingallinoneragframework}: graphrag supporting multimodal retrieval. (5) HippoRAG\cite{gutiérrez2025ragmemorynonparametriccontinual}: a neuro-inspired graph retriever traverses the graph with Personalized PageRank.

Retrieval hyperparameters are as follows: BM25 uses standard Okapi parameters (k1=1.5, b=0.75).RagAnything (Local) retrieves the top-10 seed entities with one hop expansion. RagAnything (Mix) starts with top-10 seed candidates, then expands to the top-8 neighbors per seed. HippoRAG uses its hierarchical mechanism with Personalized PageRank (damping factor=0.5) to propagate relevance and enable multi-hop traversal to relevant passages. 

\newpage
\subsection{Prompt Templates}
\label{app:prompt}
\begin{tcolorbox}[
  enhanced,
  breakable,
  colback=blue!2,
  colframe=blue!55!black,
  boxrule=0.6pt,
  arc=2.5pt,
  left=6pt,right=6pt,top=6pt,bottom=6pt,
  title=\textbf{Hierarchical Entity Extraction Prompt},
  fonttitle=\bfseries
]

\small

\textbf{Task Overview.}
You are given a chart (table, figure, or a combination of both), its caption,
and optional surrounding context.
Your task is to construct a \textbf{hierarchical knowledge graph} that represents the semantic structure of the chart.

\textbf{Overall Objective.}
Your goal is to build a \textbf{layered knowledge graph}:
\begin{itemize}
  \item \textbf{Level 1 entities} represent the chart's core topic and
  primary comparison dimensions, derived from the caption and top-level headers.
  \item \textbf{Level 2 entities} represent secondary dimensions or detailed
  subcategories, such as rows, subheaders, or nested categories.
  \item \textbf{Level 3 and beyond} represent supporting attributes at a finer
  level of granularity.
\end{itemize}

The number of levels should be \textbf{adaptively determined} by the chart
structure. If a level can be decomposed further (e.g., via multi-level headers),
you must recursively create deeper levels.

Entity Extraction Guidelines.
\begin{enumerate}
  \item Extract \textbf{semantic-rich and disambiguated entities}.
  \begin{itemize}
    \item Level 1: Theme entities and key dimensions.
    \item Level 2: Row or column entities specifying subcategories.
    \item Level 3+: Fine-grained attributes.
    \item Always include qualifiers for uniqueness (e.g., ``Revenue (2023)'').
    \item All entities must belong to the required entity types
    \texttt{\{entity\_types\}}.
  \end{itemize}

  \item Skip generic or referential phrases
  (e.g., ``this table'', ``the data'') and presentation-only text
  (e.g., ``see below'').

  \item Provide concise and meaningful descriptions for each entity.
\end{enumerate}

\textbf{Relationship Guidelines.}
\begin{enumerate}
  \item Extract \textbf{intra-level} relationships, especially among Level 1 entities.
  \item Extract \textbf{inter-level} relationships linking higher-level entities
  to their children.
  \item All relationships must belong to the required types
  \texttt{\{relation\_types\}}.
  \item Each relationship must include:
  \begin{itemize}
    \item source\_entity and target\_entity,
    \item relation\_type,
    \item relationship\_description,
    \item relationship\_keywords.
  \end{itemize}
\end{enumerate}

Output Format.
Return a JSON object with two fields:
\texttt{entities} and \texttt{relationships}.

\textbf{Additional Constraints.}
\begin{itemize}
  \item Do not extract numerical data points.
  \item Ensure all relationships refer to extracted entities.
  \item If the input chart is too short or lacks sufficient detail,
  return an empty JSON object.
\end{itemize}

Return \textbf{only valid JSON}.  
Do not include explanations, markdown, or code blocks.
The output must start with \texttt{\{} and end with \texttt{\}}.

Chart Input.
The chart image or text is provided as \texttt{\{input\_text\}}.

\end{tcolorbox}

% === 把这段代码加在 figure 开始之前 ===
\definecolor{mdblue}{RGB}{30, 90, 180}
\definecolor{mdred}{RGB}{200, 30, 30}
\definecolor{mdgreen}{RGB}{30, 150, 80}
\definecolor{boxbg}{RGB}{245, 247, 250}   % 修复 boxbg 错误
\definecolor{framecol}{RGB}{40, 60, 90}   % 修复 framecol 错误
% ===================================

\begin{figure*}[t!]
\centering
% 主容器
\begin{tcolorbox}[
    enhanced,
    breakable,
    colback=boxbg,
    colframe=framecol,
    boxrule=0.8pt,
    arc=3pt,
    left=6pt,right=6pt,top=8pt,bottom=8pt,
    title=\textbf{\large Unified Hierarchical Entity Extraction \& QA Generation Prompt},
    fonttitle=\bfseries\large,
    subtitle style={boxrule=0pt, colback=framecol!10!white}
]

% ==================== SHARED CONTEXT ====================
\textbf{\textcolor{mdblue}{SYSTEM CONTEXT:}} You are designing a high-quality multi-chart QA pair. 
\textbf{Inputs provided:} \texttt{reasoning\_paths}, \texttt{available\_sources}, \texttt{chart\_index}, \texttt{text\_evidence}.

\vspace{2pt}
\textbf{UNIVERSAL CONSTRAINTS (Apply to ALL tasks):}
\begin{itemize}[noitemsep,topsep=0pt,leftmargin=1.5em]
    \item \textbf{Self-contained:} No deictic references (e.g., "this chart"). Include scope qualifiers.
    \item \textbf{Multi-source:} Must combine evidence from at least TWO sources (Chart+Chart or Chart+Text).
    \item \textbf{\textcolor{mdred}{CRITICAL: Do NOT reveal specific numbers/percentages in the question text.}}
\end{itemize}

\vspace{4pt}\hrule\vspace{4pt}

% ==================== DIFFERENTIATED MODULES ====================
\textbf{\large Select ONE Strategy Module based on Task Type:}

\vspace{4pt}

% --- Row 1: FactCheck & Manipulation ---
\begin{tcbitemize}[raster columns=2, raster equal height, raster column skip=6pt]
    % Module A: FactCheck
    \tcbitem[colframe=mdblue!80!black, title=\textbf{Module A: QA\_FactCheck}, colback=white]
    \small
    \textbf{Goal:} Verify a claim using evidence.
    \begin{itemize}[noitemsep,leftmargin=*]
        \item \textbf{Question:} Must present a \textbf{CLAIM} (e.g., "Is it true that...").
        \item \textbf{Answer:} Must be binary ("True"/"False", "Yes"/"No") + Evidence.
        \item \textbf{Logic:} Verify truth value across sources.
    \end{itemize}
    \textit{Ex: "Did the growth rate exceed 5\% in 2023?"}
    
    % Module B: Manipulation
    \tcbitem[colframe=mdred!80!black, title=\textbf{Module B: QA\_Manipulation}, colback=white]
    \small
    \textbf{Goal:} Require arithmetic calculation.
    \begin{itemize}[noitemsep,leftmargin=*]
        \item \textbf{Question:} Ask for difference, sum, ratio, or growth rate.
        \item \textbf{Answer:} Specific numerical value derived from calculation.
        \item \textbf{Logic:} $Data_A \pm Data_B$ or $Data_A / Data_B$.
    \end{itemize}
    \textit{Ex: "What is the percentage point difference between..."}
\end{tcbitemize}

\vspace{2pt}

% --- Row 2: Complex & Analysis ---
\begin{tcbitemize}[raster columns=2, raster equal height, raster column skip=6pt]
    % Module C: Complex Reasoning (Stealth)
    \tcbitem[colframe=mdgreen!80!black, title=\textbf{Module C: QA\_Complex\_Reasoning (Stealth)}, colback=white]
    \small
    \textbf{Goal:} Low lexical overlap (Hard Retrieval).
    \begin{itemize}[noitemsep,leftmargin=*]
        \item \textbf{Requirement:} \textbf{MANDATORY PARAPHRASING}.
        \item \textbf{Forbidden:} No Entity Names (use descriptions), No Years (use relative time), No Keywords (e.g., "increase").
        \item \textbf{Logic:} Functional descriptions \& relational anchoring.
    \end{itemize}
    \textit{Ex: "Among the Southeast Asian archipelago nations..."}
    
    % Module D: Analysis & Trend
    \tcbitem[colframe=orange!80!black, title=\textbf{Module D: QA\_Analysis / QA\_Trend}, colback=white]
    \small
    \textbf{Goal:} High-level pattern recognition.
    \begin{itemize}[noitemsep,leftmargin=*]
        \item \textbf{Question:} Ask about \textbf{Logical Relationships} (correlations) or \textbf{Temporal Trends} (peaks, fluctuations).
        \item \textbf{Logic:} Synthesize separate observations into a conclusion.
        \item \textbf{Depth:} Use L2/L3 reasoning branches.
    \end{itemize}
    \textit{Ex: "How did the trend in X correlate with Y...?"}
\end{tcbitemize}

\vspace{4pt}\hrule\vspace{4pt}

% ==================== SHARED OUTPUT ====================
\textbf{UNIFIED OUTPUT FORMAT (Strict JSON):}
\begin{verbatim}
{
  "query_type": "FactCheck" | "Manipulation" | "Analysis" | "Trend",
  "question": "...",         // Follows Module constraints above
  "answer": "...",           // Includes specific numbers/calcs/verdict
  "explanation": "...",      // References reasoning edges
  "entities_used": ["..."],  // List of entity IDs
  "evidence": ["..."],       // List of chunk IDs
  "difficulty_subjective": 1-3
}
\end{verbatim}

\end{tcolorbox}
\caption{Unified Prompt Template for Multi-Chart QA Generation. The system shares a common context and output format, but branches into four distinct modules (A-D) with specific logic, constraints, and paraphrasing requirements depending on the desired query type.}
\label{fig:unified_prompt}
\end{figure*}

\newpage
\subsection{Agent Environment}
\label{app:agent}

Following the multi-turn VLM-agent training paradigm, we model visual search on the global chart KG as a partially observable Markov decision process (POMDP) $\langle \mathcal{S},\mathcal{O},\mathcal{A},\mathcal{R}\rangle$. At turn $t$, the agent receives an observation $o_t\in\mathcal{O}$ (the current KG context and retrieved evidence), generates an action $a_t\in\mathcal{A}$ to further explore the KG, and the environment returns a scalar reward $r_t=\mathcal{R}(s_t,a_t)$. The objective is to maximize the expected discounted return $\max_\theta \mathbb{E}_{\pi_\theta}\big[\sum_{t=0}^{T-1}\gamma^t r_t\big]$. In 

In the standard \emph{concatenation} rollout, the context grows with turns,
$c_t^{\text{concat}}=[\text{sys},o_0,y_0,\dots,o_t]$, which quickly exceeds the VLM context window and
induces high-variance credit assignment over an ultra-long token axis.
We instead adopt a \emph{non-concatenated} rollout: at each turn $t$ the policy conditions only on the
current prompt context
\begin{equation}
c_t = [\text{sys}, o_t],
\end{equation}
and autoregressively generates a response/action token sequence
$y_t=(y_{t,1},\dots,y_{t,K_t})$ with
\begin{equation}
\pi_\theta(y_t\mid c_t)=\prod_{j=1}^{K_t}\pi_\theta(y_{t,j}\mid c_t, y_{t,<j}).
\end{equation}
Accordingly, PPO is computed \emph{per turn} without requiring a forward pass over the entire
concatenated trajectory. Let
\begin{equation}
u_{t,j}(\theta)=
\frac{\pi_\theta(y_{t,j}\mid c_t, y_{t,<j})}
{\pi_{\text{old}}(y_{t,j}\mid c_t, y_{t,<j})},
\end{equation}
and let $m_{t,j}^{\text{act}}\in\{0,1\}$ mask response/action tokens.
The PPO objective is
\begin{equation}
J_{\text{PPO}}(\theta)=
\frac{1}{\sum_{t,j} m_{t,j}^{\text{act}}}
\sum_{t,j} m_{t,j}^{\text{act}}
\min\!\Big(
u_{t,j}(\theta)A_t,\ 
\mathrm{clip}(u_{t,j}(\theta),1-\epsilon,1+\epsilon)A_t
\Big),
\label{eq:ppo_nonconcat}
\end{equation}
where the advantage $A_t$ is defined at the \emph{turn} level (below) and broadcast to tokens in the same turn.

\paragraph{Turn-level GAE and broadcasting.}
We learn a critic $V_\phi(c_t)$ defined on the turn context $c_t$.
Given turn rewards $\{r_t\}_{t=1}^{T}$, we compute TD residuals and GAE over turns:
\begin{align}
\delta_t &= r_t + \gamma V_\phi(c_{t+1}) - V_\phi(c_t), \\
A_t &= \delta_t + \gamma\lambda A_{t+1},
\end{align}
with $V_\phi(c_{T+1})\!=\!0$ for terminal.
We then assign each response token in turn $t$ the same advantage:
\begin{equation}
A_{t,j}= m_{t,j}^{\text{act}}\,A_t.
\end{equation}

\paragraph{Value regression at turn boundaries.}
Let the turn return target be $G_t = A_t + V_\phi(c_t)$ (stop-gradient on the RHS).
We regress the critic only once per turn using a value-mask $m_t^{\text{val}}\!=\!1$:
\begin{equation}
L_V(\phi)=
\frac{1}{\sum_t m_t^{\text{val}}}\sum_t m_t^{\text{val}}
\big(V_\phi(c_t)-G_t\big)^2.
\label{eq:value_nonconcat}
\end{equation}
In implementation, $V_\phi(c_t)$ can be read from a designated anchor position (e.g., the first response token)
while the objective remains a turn-level value function.

our setting, the policy $\pi_\theta$ is parameterized by a VLM that 
This design yields stable bootstrapping targets while keeping advantage estimation aligned with turn-level interaction dynamics, mitigating the instability induced by ultra-long concatenated contexts.
\newpage
\begin{figure*}[t!]
\centering
% 主容器
\begin{tcolorbox}[
    enhanced,
    breakable,
    colback=boxbg,
    colframe=framecol,
    boxrule=0.8pt,
    arc=3pt,
    left=6pt,right=6pt,top=8pt,bottom=8pt,
    title=\textbf{\large ChartWalker Agent: Navigation \& Reasoning System Prompt},
    fonttitle=\bfseries\large,
    subtitle style={boxrule=0pt, colback=framecol!10!white}
]

% ==================== SHARED CONTEXT (SYSTEM LEVEL) ====================
\textbf{\textcolor{mdblue}{SYSTEM CONTEXT:}} You are an intelligent agent navigating a multi-modal, multi-level Knowledge Graph (MMKG).
\textbf{Goal:} Explore the graph to find the correct answer for a given query.

\vspace{2pt}
\textbf{CORE OBJECTIVES:}
\begin{enumerate}[noitemsep,topsep=0pt,leftmargin=1.5em]
    \item \textbf{Identify} required information for the query.
    \item \textbf{Navigate} to entities providing this info.
    \item \textbf{Extract} concrete facts/numerical values from evidence.
    \item \textbf{Stop} once sufficient info is collected.
\end{enumerate}
\textit{Note: Exploration actions that do not fill missing information are discouraged.}

\vspace{4pt}\hrule\vspace{4pt}

% ==================== DIFFERENTIATED MODULES ====================
\textbf{\large Interaction Phase Templates:}

\vspace{4pt}

% --- Row 1: Initialization & Termination (Start/End) ---
\begin{tcbitemize}[raster columns=2, raster equal height, raster column skip=6pt]
    % Module A: Initialization
    \tcbitem[colframe=mdblue!80!black, title=\textbf{Phase 1: Initialization (Start)}, colback=white]
    \small
    \textbf{Input:} \texttt{[Initial Graph Observation]}, \texttt{Candidate Start Entities}.
    \vspace{2pt}
    \begin{itemize}[noitemsep,leftmargin=*]
        \item \textbf{Context:} Choose an entry point.
        \item \textbf{Valid Action:} \texttt{start}
        \item \textbf{Grammar:} Must choose exactly ONE entity name from the candidate list.
    \end{itemize}
    \vspace{4pt}
    \textit{Example Output:}\\
    \texttt{<answer>start CountryX</answer>}
    
    % Module C: Termination (Placed here for layout balance)
    \tcbitem[colframe=mdred!80!black, title=\textbf{Phase 3: Termination (Stop)}, colback=white]
    \small
    \textbf{Input:} Collected evidence, Observation.
    \vspace{2pt}
    \begin{itemize}[noitemsep,leftmargin=*]
        \item \textbf{Condition:} Collected enough info OR reached max stops.
        \item \textbf{Valid Action:} \texttt{stop}
        \item \textbf{Requirement:} Output final answer text immediately after operator.
    \end{itemize}
    \vspace{4pt}
    \textit{Example Output:}\\
    \texttt{<answer>stop The answer is Paris</answer>}
\end{tcbitemize}

\vspace{2pt}

% --- Row 2: Navigation Loop (Full Width for Complexity) ---
\begin{tcolorbox}[
    colframe=mdgreen!80!black, 
    title=\textbf{Phase 2: Navigation \& Reasoning Loop (Intermediate Steps)}, 
    colback=white,
    boxrule=0.8pt, arc=2pt
]
\small
\textbf{Input:} \texttt{[Current Graph State]} (Entity desc, Relations, Memory, Visited sources).
\par\vspace{2pt}
\textbf{Reasoning:} Decide next action based on context/history. \textbf{\textcolor{mdred}{WARNING: DO NOT REPEAT ACTIONS OR SEARCH FORBIDDEN RELATIONS.}}

\vspace{4pt}
\textbf{Action Grammar Options:}
\begin{itemize}[noitemsep,topsep=2pt,leftmargin=1.5em]
    \item \textbf{\texttt{edge\_search <int>}}: Inspect a relation index from \texttt{[Searchable Relations]}.\\
    \textit{Ex: \texttt{<answer>edge\_search 1</answer>}}
    
    \item \textbf{\texttt{move <int>}}: Move to a relation's target entity (to get more info). Targets can be from Searchable OR Forbidden lists.\\
    \textit{Ex: \texttt{<answer>move 1</answer>}}
    
    \item \textbf{\texttt{backward <Name>}}: Return to a previously visited entity.\\
    \textit{Ex: \texttt{<answer>backward CountryY</answer>}}
\end{itemize}
\end{tcolorbox}

\vspace{4pt}\hrule\vspace{4pt}

% ==================== OUTPUT FORMAT ====================
\textbf{UNIFIED OUTPUT FORMAT:} All responses must be wrapped in XML-style tags.
\begin{verbatim}
<answer> [action_keyword] [arguments] </answer>
\end{verbatim}

\end{tcolorbox}
\caption{The ChartWalker Agent Prompt Structure. The prompt guides the agent through three distinct phases: (1) Selecting a start entity, (2) An iterative navigation loop involving edge searching and entity traversal, and (3) A termination phase to output the final answer.}
\label{fig:mmkg_prompt}
\end{figure*}

\newpage
\subsection{Showcase}
\label{app:show_case}
\begin{figure}[!ht]
\centering
\begin{tcolorbox}[chartcard, width=0.85\linewidth]

{\bfseries \# Manipulation\par\vspace{6pt}}

% -----------------------------
% KG path
% -----------------------------
{\bfseries \#\# KG Path}\par
{\color{mdred}
Democrats/leaning Democrats (L1) --horizontal [src=chart\_00189\_07]--> Global Climate Change (L1)\par
Global Climate Change (L1) --vertical\_up [src=chart\_00048\_01]--> Survey (L0)\par
Democrats/leaning Democrats (L1) --vertical\_down [src=chart\_00189\_07]--> Catholic Democrats\par
}\par

% -----------------------------
% Query
% -----------------------------
{\bfseries \#\# Query}\par
{\color{mdblue}
In 2022, what percentage of U.S. Catholic adults who identify as Democrats or leaning Democratic
viewed global climate change as an extremely or very serious problem, and how does this compare
to the percentage of all U.S. adults who identify as Democrats or leaning Democratic who considered
dealing with global climate change a top priority for the president and Congress in 2024?
}\par\vspace{10pt}

% -----------------------------
% Answer
% -----------------------------
{\bfseries \#\# Answer}\par
{\color{mdblue}
82\% in 2022 among Catholic Democrats, which is 23 percentage points higher than the 59\% in 2024 among all Democrats.
}\par\vspace{10pt}

% -----------------------------
% Evidence
% -----------------------------
{\bfseries \#\# Evidence}\par

% -----------------------------
% Two charts side-by-side
% -----------------------------
\begin{center}
\begin{minipage}{0.48\linewidth}
\centering
{\small \color{mdgreen}chart\_00189\_07 (2022, Catholic Dem/lean Dem)}\par\vspace{4pt}
\begin{overpic}[width=\linewidth]{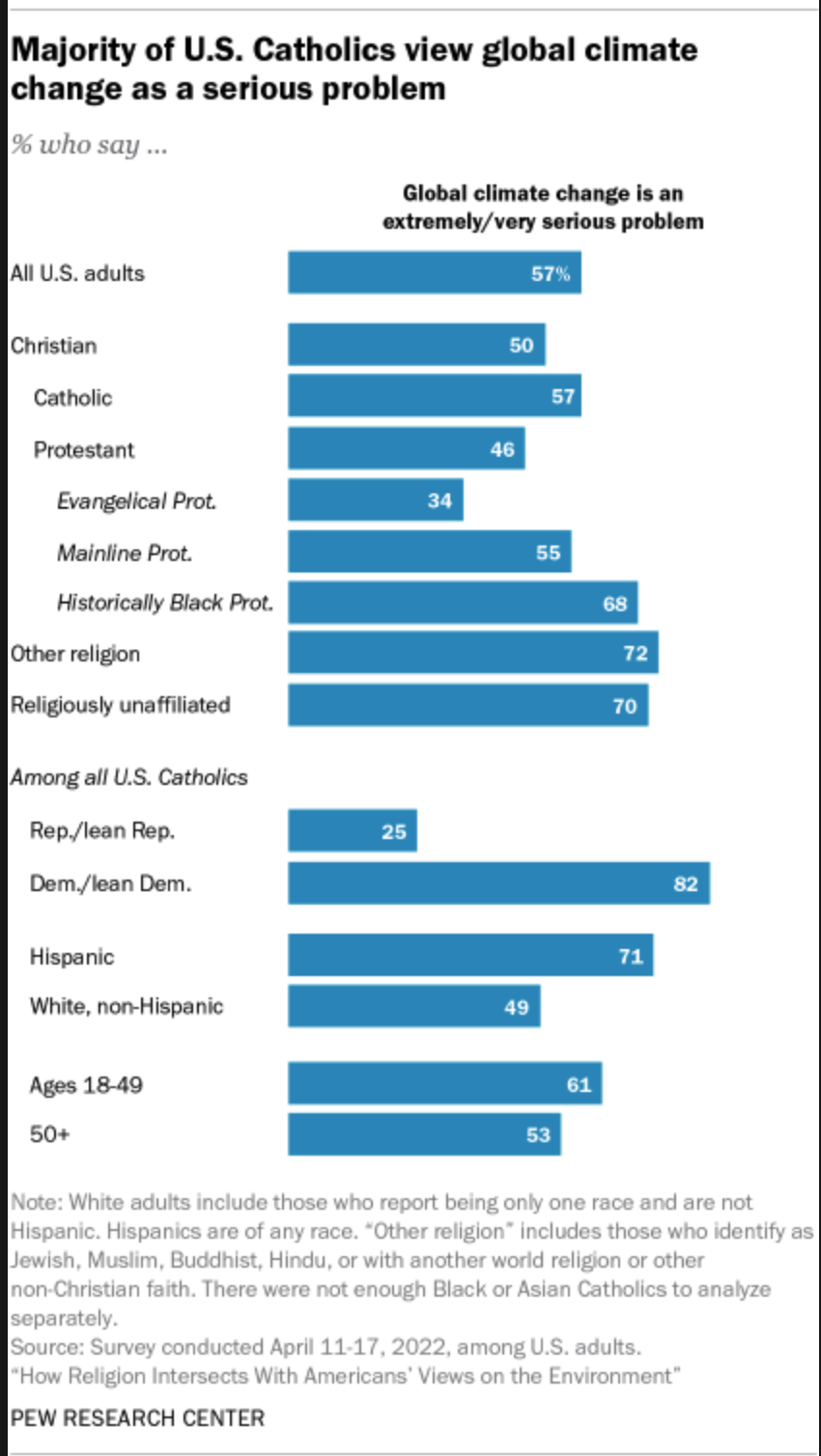}
\end{overpic}
\end{minipage}
\hfill
\begin{minipage}{0.48\linewidth}
\centering
{\small \color{mdgreen}chart\_00048\_01 (2024, all Dem/lean Dem)}\par\vspace{4pt}
\begin{overpic}[width=\linewidth]{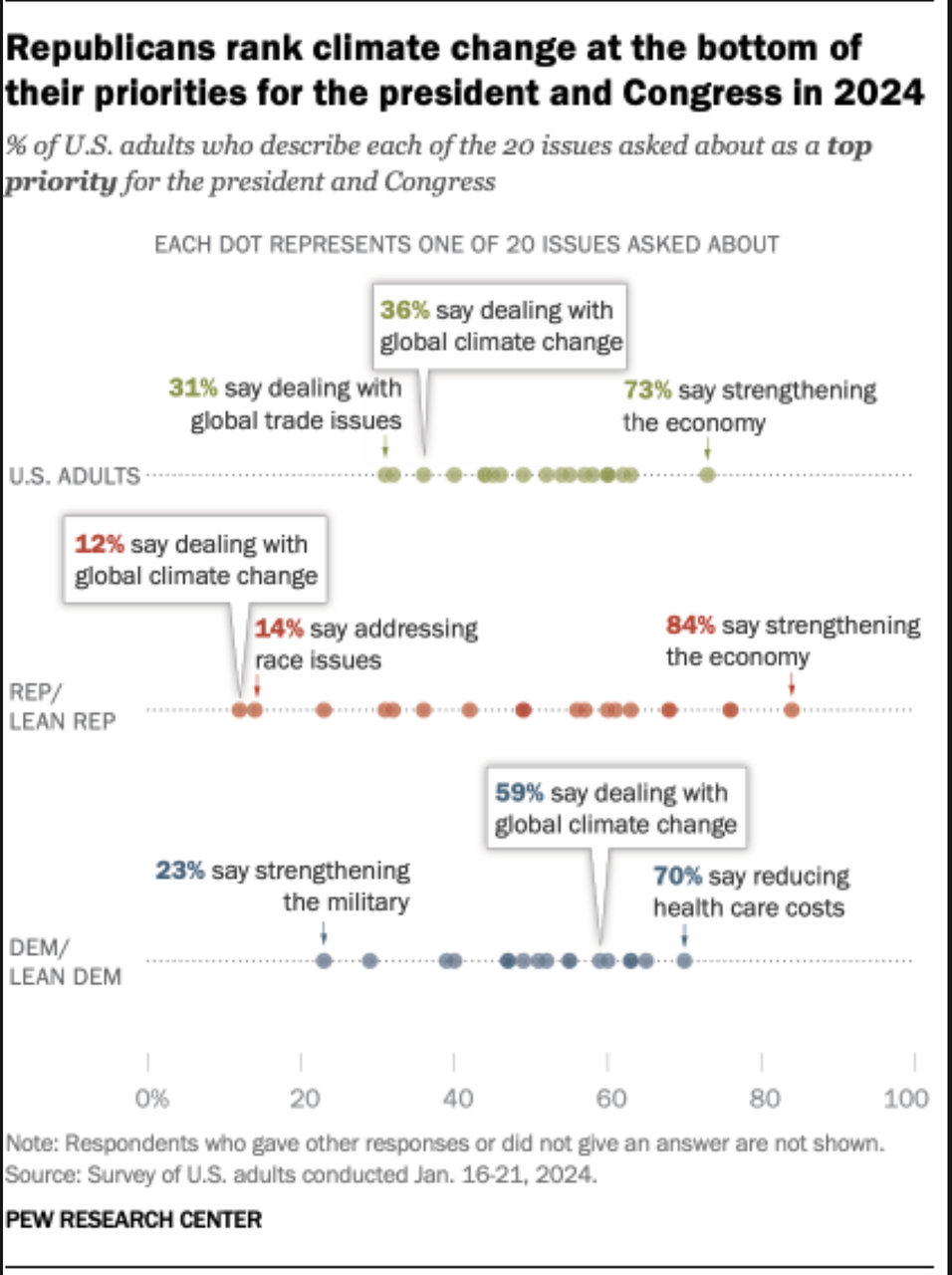}
\end{overpic}
\end{minipage}
\end{center}

\vspace{8pt}

\end{tcolorbox}

\caption{Manipulation showcase. Due to the relation of Democrats and Catholic Democrats being highlighted in the sampled path, our problem does not contain the logical inconsistency shown in \cite{lu2025deepdiveadvancingdeepsearch} (Figure~\ref{fig:wrong_image}).}
\end{figure}

\begin{figure}
    \centering
    \includegraphics[width=0.9\linewidth]{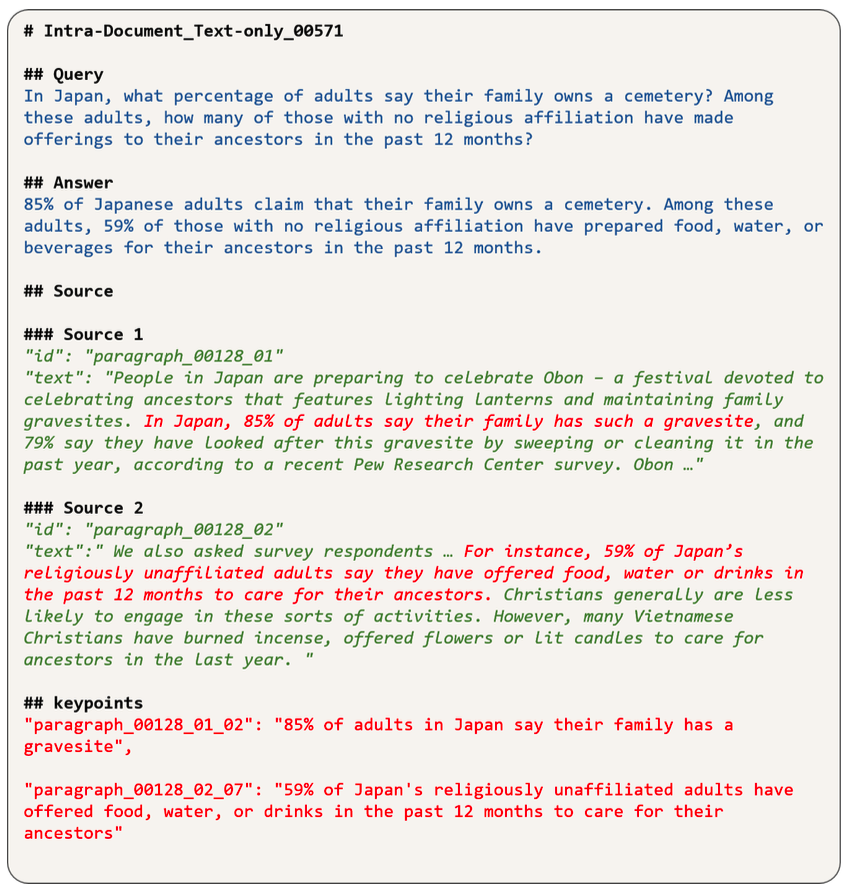}
    \caption{Wrong case}
    \label{fig:wrong_image}
\end{figure}

\begin{figure}[t]
\centering
\begin{tcolorbox}[chartcard, width=0.85\linewidth]

{\bfseries \# Complex\_Reasoning\par\vspace{6pt}}

% -----------------------------
% KG path
% -----------------------------
{\bfseries \#\# KG Path}\par
{\color{mdred}
USA (L1) --horizontal [src=test\_000774]--> Coal (33\%)\par
USA (L1) --horizontal [src=test\_000774]--> Natural gas (33\%)\par
SocialMediaPlatform (L1) --horizontal [src=test\_001503]--> Clean \& renewable energy (35\%)\par
}
% Top2(USA electricity generation, 2015) = \{Coal, Natural gas\}\par
% Top1(platform ops energy, 2015) = \{Clean \& renewable energy\}\par
% $\text{Top2} \cap \text{Top1} = \emptyset \;\Rightarrow\; P_{\text{top1}} \notin G_{\text{top2}}$\par
% }\par

% -----------------------------
% Query
% -----------------------------
{\bfseries \#\# Query}\par
{\color{mdblue}
For the same year (2015), let $G_{\text{top2}}$ be the set of the top-2 energy sources by share in the national
electricity generation mix, and let $P_{\text{top1}}$ be the top-1 energy source by share in the social media
platform’s operations energy consumption breakdown. Check whether $P_{\text{top1}} \in G_{\text{top2}}$.
}\par\vspace{10pt}

% -----------------------------
% Answer
% -----------------------------
{\bfseries \#\# Answer}\par
{\color{mdblue}
No. In 2015, $G_{\text{top2}}=\{\text{Coal},\ \text{Natural gas}\}$, while $P_{\text{top1}}=\text{Clean \& renewable energy}$, so $P_{\text{top1}} \notin G_{\text{top2}}$.
}\par\vspace{10pt}

% -----------------------------
% Evidence
% -----------------------------
{\bfseries \#\# Evidence}\par

% -----------------------------
% Charts
% -----------------------------
\begin{center}
\begin{minipage}{0.48\linewidth}
\centering
{\small \color{mdgreen}test\_000774 (2015, USA electricity generation mix)}\par\vspace{4pt}
\begin{overpic}[width=\linewidth]{image/test\_000774.png}
\end{overpic}
\end{minipage}
\hfill
% \begin{minipage}{0.48\linewidth}
% \centering
% {\small \color{mdgreen}test\_001503 (2015, platform ops energy breakdown)}\par\vspace{4pt}
% \begin{overpic}[width=\linewidth]{image/test\_001503.png}
% \end{overpic}
% \end{minipage}

% \vspace{6pt}

\begin{minipage}{0.60\linewidth}
\centering
{\small \color{mdgreen}test\_000699 (supporting chart / context)}\par\vspace{4pt}
\begin{overpic}[width=\linewidth]{image/test\_000699.png}
\end{overpic}
\end{minipage}
\end{center}

\vspace{8pt}

\end{tcolorbox}

\caption{Complex Reasoning showcase.}
\end{figure}

\begin{figure}[t]
\centering
\begin{tcolorbox}[chartcard, width=0.85\linewidth]

{\bfseries \# FactCheck\par\vspace{6pt}}

% -----------------------------
% -----------------------------
% KG path
% -----------------------------
{\bfseries \#\# KG Path}\par
{\color{mdred}
Peru (L1) --horizontal [src=chart\_00260\_02]--> Favorable opinion of Brazil (58\%)\par
Peru (L1) --horizontal [src=chart\_00260\_04]--> Confidence in Brazil's president (30\%)\par
Spring 2024 Pew surveys (L1) --horizontal [src=paragraph\_00260\_01]--> Covered countries include Peru\par
}\par

% -----------------------------
% Query
% -----------------------------
{\bfseries \#\# Query}\par
{\color{mdblue}
In the spring 2024 surveys, does the share of people in Peru with a favorable opinion of Brazil exceed
the share expressing confidence in Brazil's president?
}\par\vspace{10pt}

% -----------------------------
% Answer
% -----------------------------
{\bfseries \#\# Answer}\par
{\color{mdblue}
Yes (58\% vs 30\%).
}\par\vspace{10pt}

% -----------------------------
% Two charts side-by-side
% -----------------------------
\begin{center}
\begin{minipage}{0.48\linewidth}
\centering
{\small \color{mdgreen}chart\_00260\_02 (Peru favorable opinion of Brazil, Spring 2024)}\par\vspace{4pt}
\begin{overpic}[width=\linewidth]{image/chart\_00260\_02_fixed.png}
\end{overpic}
\end{minipage}
\hfill
\begin{minipage}{0.48\linewidth}
\centering
{\small \color{mdgreen}chart\_00261\_04 (Peru confidence in Lula, Spring 2024)}\par\vspace{4pt}
\begin{overpic}[width=\linewidth]{image/chart\_00261\_04_fixed.png}
\end{overpic}
\end{minipage}
\end{center}

\vspace{8pt}

\end{tcolorbox}

\caption{Factcheck showcase.}
\end{figure}

\begin{figure}[t]
\centering
\begin{tcolorbox}[chartcard, width=0.85\linewidth]

{\bfseries \# Analysis\par\vspace{6pt}}

% -----------------------------
% KG path
% -----------------------------
{\bfseries \#\# KG Path}\par
{\color{mdred}
Israel (L1) --horizontal [src=chart\_00076\_03]--> EU favorability, Ideological left (85\%)\par
Israel (L1) --horizontal [src=chart\_00076\_03]--> EU favorability, Ideological right (49\%)\par
Israel (L1) --horizontal [src=chart\_00235\_04]--> Biden approval/confidence, Ideological left (48\%)\par
Israel (L1) --horizontal [src=chart\_00235\_04]--> Biden approval/confidence, Ideological right (61\%)\par
}\par

% -----------------------------
% Query
% -----------------------------
{\bfseries \#\# Query}\par
{\color{mdblue}
Among Israeli adults, does a more favorable view of the European Union correlate with lower approval of U.S. President Biden?
}\par\vspace{10pt}

% -----------------------------
% Answer
% -----------------------------
{\bfseries \#\# Answer}\par
{\color{mdblue}
Yes — Israeli adults on the ideological left show higher EU favorability (85\%) but lower Biden approval/confidence (48\%),
while the ideological right shows lower EU favorability (49\%) but higher Biden approval/confidence (61\%).
}\par\vspace{10pt}

% -----------------------------
% Evidence
% -----------------------------
{\bfseries \#\# Evidence}\par

% -----------------------------
% Two charts side-by-side
% -----------------------------
\begin{center}
\begin{minipage}{0.48\linewidth}
\centering
{\small \color{mdgreen}chart\_00076\_03 (Israel, EU favorability by ideology)}\par\vspace{4pt}
\begin{overpic}[width=\linewidth]{image/chart\_00076\_03_fixed.png}
\end{overpic}
\end{minipage}
\hfill
\begin{minipage}{0.48\linewidth}
\centering
{\small \color{mdgreen}chart\_00235\_04 (Israel, Biden approval/confidence by ideology)}\par\vspace{4pt}
\begin{overpic}[width=\linewidth]{image/chart\_00235\_04\_fixed.png}
\end{overpic}
\end{minipage}
\end{center}

\vspace{8pt}

\end{tcolorbox}

\caption{Analysis showcase}
\end{figure}